\documentclass[saare,prev, superscriptaddress, preprintnumbers,floatset,nofootinbib,twocolumn]{revtex4-1}
\pdfoutput=1
\usepackage[english]{babel}
\usepackage{bm}
\usepackage{times}
\usepackage{ulem}
\usepackage{hyperref}
\hypersetup{
    colorlinks=true,
    linkcolor=blue,
    filecolor=magenta,      
    urlcolor=blue,
    citecolor=blue,
    pdftitle={Sharelatex Example},
    pdfpagemode=FullScreen
    }
\usepackage{listings}
\usepackage{slashed}
\usepackage{color}
\usepackage{appendix}
\usepackage{mathtools}
\usepackage{epsfig}
\usepackage{dcolumn}
\usepackage{textcomp}
\usepackage{multirow}
\usepackage{graphicx}
\usepackage{soul}
\usepackage{subfigure}

\newcommand{\fig}[1]{Fig.~\ref{#1}} 
 
\newcommand{\eq}[1]{Eq.~(\ref{#1})}

\begin{document}

\title{Probing a Dark Sector with Collider Physics, Direct Detection, and Gravitational Waves}

\author{Giorgio Arcadi}
\email{giorgio.arcadi@unime.it}
\affiliation{Dipartimento di Scienze Matematiche e Informatiche, Scienze Fisiche e Scienze della Terra, Universita degli Studi di Messina, Via Ferdinando Stagno d’Alcontres 31, I-98166 Messina, Italy}
\affiliation{INFN Sezione di Catania, Via Santa Sofia 64, I-95123 Catania, Italy}

\author{Glauber C. Dorsch}
\email{glauber@fisica.ufmg.br}
\affiliation{Departamento de F\'isica, Universidade Federal de Minas Gerais, 31270-901, Belo Horizonte, MG, Brasil}

\author{Jacinto P. Neto}
\email{jacinto.neto.100@ufrn.edu.br}
\affiliation{Departamento de F\'isica, Universidade Federal do Rio Grande do Norte, 59078-970, Natal, RN, Brasil}
\affiliation{International Institute of Physics, Universidade Federal do Rio Grande do Norte,  59078-970, Natal, RN, Brasil}

\author{Farinaldo S. Queiroz}
\email{farinaldo.queiroz@ufrn.br}
\affiliation{Departamento de F\'isica, Universidade Federal do Rio Grande do Norte, 59078-970, Natal, RN, Brasil}
\affiliation{International Institute of Physics, Universidade Federal do Rio Grande do Norte,  59078-970, Natal, RN, Brasil}
\affiliation{Millennium Institute for Subatomic Physics at High-Energy Frontier (SAPHIR), Fernandez Concha 700, Santiago, Chile}

\author{Y. M. Oviedo-Torres}
\email{ymot@estudantes.ufpb.br}
\affiliation{International Institute of Physics, Universidade Federal do Rio Grande do Norte,  59078-970, Natal, RN, Brasil}
\affiliation{Departamento de F\'isica, Universidade Federal da Para\'iba, 58051-970, Jo\~ao Pessoa, PB, Brasil}

\begin{abstract}
We assess the complementarity between colliders, direct detection searches, and gravitational wave interferometry in probing a scenario of dark matter in the early universe. The model under consideration contains a $B-L$ gauge symmetry and a vector-like fermion which acts as the dark matter candidate. The fermion induces significant a large dark matter-nucleon scattering rate, and the $Z^\prime$ field produces clear dilepton events at colliders. Thus, direct detection experiments and colliders severely constrain the parameter space in which the correct relic density is found in agreement with the data. Nevertheless, little is known about the new scalar responsible for breaking the B-L symmetry. If this breaking occurs via a first-order phase transition at a TeV scale, it could lead to gravitational waves in the mHz frequency range detectable by LISA, DECIGO, and BBO instruments. The spectrum is highly sensitive to properties of the scalar sector and gauge coupling. We show that a possible GW detection, together with information from colliders and direct detection experiments, can simultaneously pinpoint the scalar self-coupling, and narrow down the dark matter mass where a thermal relic is viable.

\noindent

\end{abstract}

\keywords{}

\maketitle
\flushbottom

\section{\label{sec:intro} Introduction}

The evidence for the presence of dark matter (DM) in the universe is compelling, and thermal relics stand out among the possible explanations for it. They typically experience interactions that could be probed by current and near-future experiments and easily yield the correct relic density, thus attracting much attention from the community. The nature of these particles allowed us to explore the so-called dark matter complementarity, which refers to the use of data from various sources, such as direct and indirect detection experiments, as well as colliders, as a way to narrow down the viable properties of a dark matter particle in a given model (see \cite{Arcadi:2017kky} for a review). With the detection of gravitational waves (GWs) by LIGO/Virgo/KAGRA~\cite{LIGOScientific:2016aoc, LIGOScientific:2018mvr, LIGOScientific:2021usb, LIGOScientific:2021djp}, together with the NANOGrav evidence for a stochastic GW background~\cite{NANOGrav:2023gor}, and the near-future launch of new-generation space-based interferometers~\cite{LISA:2017pwj}, a new era has surfaced. 
If the scalar in the dark sector was responsible for a first-order phase transition, there could be a resulting a stochastic gravitational wave spectrum which could indirectly constrain the dark sector. For 
a DM model at TeV scale, the spectrum would be in the range of optimal detectability at LISA/DECIGO/BBO~\cite{Crowder:2005nr, Kawamura:2006up}. In other words, gravitational waves have become an interesting laboratory for dark sectors that feature a scalar particle inducing a first-order phase transition.

In this paper, we illustrate the \textit{complementarity} between colliders, direct detection experiments, and gravitational wave detectors for testing models with a dark sector. Our main argument is that any near future collider will likely be insensitive to the self-couplings of a dark scalar, but this parameter could be constrained from possible detections of cosmological GWs induced from this scalar sector in a first-order phase transition. On the other hand, the GW spectrum depends also on the gauge and fermionic sectors, so GW detectors alone would not be able to disentangle the information on the self-coupling. More concretely, we will show, in a specific model, that none of these experiments alone could probe all dimensions of the parameter space, but a full picture could be obtained by considering the symbiosis of colliders, detectors, and GW experiments. 

To be concrete, we work on a model with an additional $U(1)_{B-L}$ gauge boson $Z^\prime$, together with a DM candidate, which is a Dirac vector-like fermion coupled to the visible sector via a $Z^\prime$ portal. By computing the spin-independent nucleon-DM scattering cross-section, we can impose constraints on the model due to direct detection limits from XENONnT and LUX-ZEPLIN (LZ) \cite{XENON:2023sxq, LZ:2022ufs}. Moreover, collider searches by ATLAS and LEP II impose bounds on the mass and coupling of the $Z^\prime$ \cite{ATLAS:2019erb,ATLAS:2019fgd,ATLAS:2019npw}. But these experiments can say little about the scalar sector responsible for breaking the $U(1)_{B-L}$ symmetry: the scalar is not relevant for the DM phenomenology and may decay only into invisible particles, making it difficult to detect at colliders.  
However, the $U(1)_{B-L}$ symmetry-breaking process may be a first-order phase transition in the early Universe, resulting in GWs testable at the aforementioned detectors, as shown in Ref.~\cite{Hasegawa:2019amx,Bosch:2023spa}. The GW spectrum is highly sensitive to the scalar sector but depends also on the $Z^\prime$ and dark matter couplings. Thus, the parameters can only be disentangled with the aid of direct detection and collider experiments. We show that, once we have a measurement of these couplings, a detection of cosmological GWs could even lead to a measurement of the scalar self-coupling at good precision using GWs. This illustrates how GW detectors, collider, and direct detection searches are complementary probes for dark sectors and beyond the Standard Model (SM) physics in general.

The DM phenomenology has been explored in the context of the minimal $B-L$ model. In \cite{Okada:2016gsh}, the authors contrasted their theoretical results with the bounds from collider and direct detection on a minimal $B-L$ model in which the lightest right-handed neutrino is the DM candidate. In \cite{Rodejohann:2015lca}, they look into the DM phenomenology of a scalar singlet dark matter charged under $B-L$, where gauge and scalar interactions are considered. Moreover, for the latter case, semi-annihilation processes are very important regarding the relic density. 

The discussion of GWs in the context of DM models is not new. Ref.~\cite{Madge:2018gfl, Breitbach:2018ddu} have explored GWs working in a secluded DM model and in a gauged $U(1)_{\ell}$ leptonic Abelian symmetry extension, respectively. In \cite{Abe:2023zja}, the authors studied a $SU(2)_{0}\times SU(2)_{1} \times SU(2)_{2}$ gauge extension. Ref.~\cite{Costa:2022lpy,Costa:2022oaa} investigated the DM phenomenology and GW production in two-component DM models. In \cite{Beniwal:2017eik, Arcadi:2022lpp} the authors worked in a similar vein for scalar sector extension models. 

One could wonder that our paper is a combination of Ref.~\cite{Okada:2016gsh} and Ref.~\cite{Hasegawa:2019amx}, however, our work differs from others because our dark matter candidate is taken to be a vector-like fermion, and new collider bounds are derived and updated limits from direct detection experiments are employed. Moreover, we generated the GW spectra for choices of the gauge coupling and the scalar singlet self-coupling, bearing in mind the parameter space favored by the dark matter phenomenology. Thus, exploiting the interplay between dark matter and GW physics. Our emphasis is not on showing that the model could yield a detectable GW spectrum, but especially on highlighting the complementarity between GW detectors, collider searches, and direct detection experiments for probing the phenomenology of a model with a gauged $B-L$ symmetry, opening also a portal-like to the scalar sector. This illustrates that the inclusion of GW detectors in the particle physicist's experimental arsenal will \textit{not} render collider and direct detection experiments obsolete, but will only enrich their results. 
 
This work is organized as follows: In Section \ref{sec:model}, we present the minimal $B-L$ model describing the particle content and the properties of the scalar singlet and DM candidate that we are interested in. The generation of GWs via first-order phase transition in the context of the minimal $B-L$ model is discussed in Section \ref{sec:gwBL}, and the DM production in Section \ref{sec:dmproduction}. We explain the DM model constraints in Section \ref{sec:bounds}. Finally, we discuss the results and present the conclusions in Sections \ref{sec:results} and \ref{sec:conclusions}, respectively.

\section{\label{sec:model} The Minimal B-L Model}
The minimal $B-L$ model is a simple extension of the Standard Model (SM). An additional $U(1)_{B-L}$ Abelian symmetry enlarges the SM gauge structure to $SU(3)_{C}\otimes SU(2)_{L} \otimes U(1)_{Y} \otimes U(1)_{B-L}$, where $B$ stands for baryon number and $L$ for lepton number. Hence, the gauge content increases by one new gauge boson, say, $Z^\prime$. In the fermion sector, we add three Majorana right-handed neutrinos $N_{iR}$ ($i=1,2,3)$ to cancel out the gauge anomalies, and the DM candidate is a vector-like Dirac fermion $\chi$, which also carries $B-L$ charge as the other fermions. Consequently, the DM-SM interactions happen through a $Z^\prime$ portal. Moreover, we include a scalar singlet $\Phi_s$, which breaks the $B-L$ symmetry and gives rise to a Majorana mass term for right-handed neutrinos that is essential to realize the seesaw mechanism type I to turn active neutrinos to massive particles. In Table~\ref{tab:particlecontent}, we present the matter particle content and their respective charge assignments under each symmetry, including the $Z_2$, which arises after $B-L$ breaking and ensures DM stability \cite{Klasen:2016qux}.

\begin{table}
\begin{tabular}{|c | c c c c c|} 
 \hline 
 \hline
  & $SU(3)_{C}$ & $SU(2)_L$ & $U(1)_Y$ & $U(1)_{B-L}$ & $Z_{2}$\\ 
 \hline
 $q_{iL}$ & $\boldsymbol{3}$ & $\boldsymbol{2}$ & $1/6$ & $1/3$ & $+$\\ 
 $u_{iR}$ & $\boldsymbol{3}$ & $\boldsymbol{1}$ & $2/3$ & $1/3$ & $+$ \\
 $d_{iR}$ & $\boldsymbol{3}$ & $\boldsymbol{1}$ & $-1/3$ & $1/3$ & $+$\\
 $\ell_{iL}$ & $\boldsymbol{1}$ & $\boldsymbol{2}$ & $-1/2$ & $-1$ & $+$ \\
 $e_{iR}$ & $\boldsymbol{1}$ & $\boldsymbol{1}$ & $-1$ & $-1$ & $+$\\ 
 $N_{iR}$ & $\boldsymbol{1}$ & $\boldsymbol{1}$ & $0$ & $-1$& $+$ \\
$\chi$ & $\boldsymbol{1}$ & $\boldsymbol{1}$ & $0$ & $1/3$& $-$ \\
 $H$  & $\boldsymbol{1}$ & $\boldsymbol{2}$ & $-1/2$ & $0$& $+$ \\
 $\Phi_s$  & $\boldsymbol{1}$ & $\boldsymbol{1}$ & $0$ & $2$& $+$ \\
 \hline \hline
\end{tabular}
\caption{\label{tab:particlecontent} The matter particle content of the minimal B-L model and their respective charge assignments under each symmetry.}
\end{table}

The Lagrangian that describes the DM phenomenology and encodes the process of $B-L$ symmetry breaking can be written as  
\begin{align}
    \mathcal{L} &\supset i\overline{\chi}\gamma^{\mu}\partial_\mu\chi - m_\chi \overline{\chi}\chi \nonumber\\
    &- \frac{1}{4}F^{\prime\,\mu\nu}F^\prime_{\mu\nu} +  g_{BL}n_\chi\overline{\chi}\gamma^{\mu}\chi Z^{\prime}_\mu \nonumber \\
    &+g_{BL}n_\ell \sum_{\psi = \ell, \nu_{\ell}} \overline{\psi}\gamma^{\mu}\psi Z^{\prime}_{\mu} +g_{BL}n_q \sum_{i = 1}^{6} \overline{q}_i\gamma^{\mu}q_i Z^{\prime}_{\mu} \nonumber \\
    & + y_{ij}^{D}\overline{L}_i \Tilde{H}N_{jR} + y_{ij}^{M} \overline{(N^{C}_{iR})}\Phi_s N_{jR} \nonumber\\
    &+ (D_{\mu}\Phi_s)^{\dagger}(D^{\mu}\Phi_s) - V_0(\Phi_s),\label{eq:lagrangianmodel}
\end{align}
where $F^{\prime}_{\mu\nu}$ and $g_{BL}$ are the strength tensor and coupling of the $B-L$ symmetry. The $n_{j}$ stands for the $B-L$ quantum number of the particles $j = \chi, \ell, \nu_\ell,$ and $q$, with $\ell = e, \mu, \tau$, and $q = u, d, c, s, t,$ and $d$. We have neglected the kinetic mixing between the photon and the $Z^\prime$ boson in our study. This is well justified because the gauge coupling $g_{BL}$ in our study will be sizeable, whereas the bounds on the kinetic mixing impose it to be suppressed, say, as strong as $10^{-2}$, yielding no effect on our phenomenology \cite{Arcadi:2017kky}. This limit was essentially due to the previous generation of $1$-Ton Direct Detection facilities as XENON1T and, consequently, it is expected to be even stronger nowadays \cite{Camargo:2018klg}. Thus, the kinetic mixing can be safely neglected throughout.

We remark that the DM charge, $n_\chi$, must be other than $\pm 1$ to avoid DM decay via an additional Yukawa term involving $\chi_R$. Notice that $\Tilde{H} = i\sigma_2 H$ is the isospin transformation of the SM Higgs doublet $H = \left(\phi^+ , \phi^0\right)^{T}$. Moreover, in the scalar sector, we have the kinetic term of $\Phi_s$ which will give mass to $Z^\prime$, and the scalar singlet potential at tree level $V_0(\Phi) =\mu_{s}^2\Phi^{\dagger}_s \Phi_s + \lambda_s \left(\Phi^\dagger_s \Phi_s \right)^2/2$. We leave further details about the effective potential $V_{\textrm{eff}}(\Phi_s)$ that leads to the first-order phase transition for the next section.

The parameterization of the scalar field singlet is given by, 
\begin{equation}
    \Phi_s = \frac{1}{\sqrt{2}}\left( v_s + \phi_s + i\rho_s \right),
\end{equation}
where $v_s = \sqrt{2} \langle \Phi_s \rangle$ is the $U(1)_{B-L}$ vacuum expectation value (VEV), and $\rho_s$ is the Goldstone boson which will be eaten by the $Z^\prime$ field after spontaneous symmetry breaking. Its mass arises from the kinetic term of scalar singlet, and the right-handed neutrinos $N_{iR}$ via the Majorana mass term in Eq.~\eqref{eq:lagrangianmodel},
\begin{align}
    m_{N_{iR}} &= \frac{y^{M}_{i}}{\sqrt{2}}v_s, \\
    m_{Z^\prime} &= 2g_{BL} v_s \label{eq:zpmasseq}.
\end{align}
We highlight that it happens at high-energy scales, say, $v_s \gg v$, where $v$ is the VEV of the SM Higgs field, $H$. In the same way, the tree-level mass of the scalar singlet is given by
\begin{equation}
    m_{\phi_s} = \sqrt{\lambda_s}v_s.
    \label{eq:ms}
\end{equation}
The active neutrinos become massive via the popular type I seesaw mechanism, which nicely reproduces the neutrino data \cite{Minkowski:1977sc,Mohapatra:1979ia,Brdar:2019iem}. 

\begin{figure*}
    \centering
    \includegraphics[width=1\linewidth]{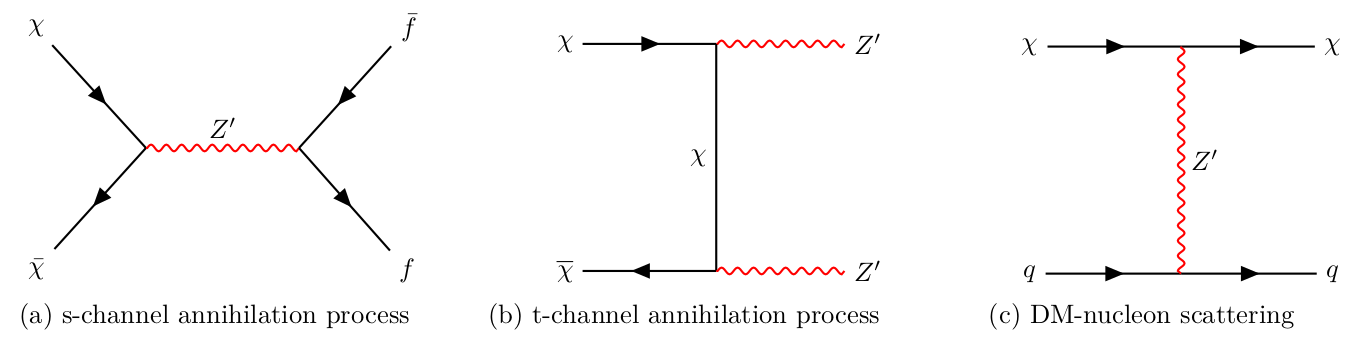}
    \caption{The main channels for DM-SM interactions in this $B-L$ DM model: (a) DM annihilation into SM fermions $f$ $\left(m_\chi < m_{Z^\prime}\right)$; (b) DM annihilation into on-shell pair of $Z^\prime$ $\left(m_\chi > m_{Z^\prime}\right)$; and (c) DM-nucleon scattering process, where $q$ stands for quarks.}
    \label{fig:feynmandigrams}
\end{figure*}

In Fig.~\ref{fig:feynmandigrams}, we display the relevant Feynman diagrams for DM phenomenology. The first and the second diagrams give the DM relic abundance. In the regime $m_\chi < m_{Z'}$, the s-channel overcomes the t-channel, and the cross-section is, 
\begin{align}
    \langle \sigma v \rangle_{\chi\overline{\chi} \rightarrow f \overline{f}} = &\frac{g^4_{BL}n^2_{\chi}n^2_{f}}{2\pi} \nonumber\\
    & \times \sum_{f}n_{c}^\textrm{f} \frac{\left( m_f^2 + 2m_{\chi}^2\right) \sqrt{1 - \frac{m^2_f}{m_\chi^2}}}{\left( m_{Z'}^2 - 4m_\chi^2\right)^2 + \Gamma^2_{Z'}m_{Z'}^2} + O(v^2),
\end{align}

where $n_f$ stands for the $B-L$ charge of the SM fermion, and $n^\textrm{f}_c$ is the number of colors of the final state SM fermion. However, when $m_\chi > m_{Z'}$, the second diagram also contributes to the DM relic abundance  whose annihilation cross-section is given by,
\begin{align}
        \langle \sigma v \rangle_{\chi\overline{\chi} \rightarrow Z' Z'} = \frac{g^4_{BL}n^4_{\chi}\left(m_{\chi}^2 - m_{Z'}^2  \right)^{3/2}}{4\pi m_{\chi}  \left( m_{Z'}^2 - 2m^2_{\chi}\right)^2} + O(v^2).
\end{align}
In summary, the free parameters that govern the DM phenomenology are the DM mass $m_\chi$, the  $Z^\prime$ mass $m_{Z^\prime}$, and the $B-L$ coupling $g_{BL}$. Concerning the third diagram in Fig.\ref{fig:feynmandigrams}, it is relevant for direct detection, as we will see later. Having reviewed the key ingredients for the relic density, in the next section, we assess the GWs production in our model due to a first-order phase transition, where $g_{BL}$ and $v_s$ play a crucial role.

\section{\label{sec:gwBL} Stochastic GW spectrum from first-order phase transition}

If the $U(1)_{B-L}$ breaking is a first-order phase transition, there are regions of broken phase in the plasma where the symmetry remains unbroken. These so-called bubbles expand and induce plasmatic motion in the form of sound waves and turbulence. At the end of the transition, when the bubbles collide and fill the entire space,  GWs are produced due to a time-varying quadrupole moment in the kinetic energy-momentum of the plasma~\cite{Caprini:2018mtu, Caprini:2019egz}. 

To estimate the shape of this spectrum, we need to study the dynamics of the phase transition. In the presence of a thermal plasma, the dynamics of the scalar field $\phi_s$ is described by an effective potential which, at 1-loop order, takes the form
\begin{equation}\begin{split}
    V_\text{eff}&(T, \phi_s)  = \dfrac{\lambda_s}{8}(\phi_s^2 - v_s^2)^2 \\
    &+ \dfrac{3\times (2g_{BL})^4}{64\pi^2}\left[ \phi_s^4 \left(\log\dfrac{\phi_s^2}{v_s^2}-\dfrac{3}{2}\right) + 2\phi_s^2v_s^2\right]\\
    &+ V_\text{th}(T,\phi_s) + \dfrac{T}{12\pi}\left(m_{Z^\prime}^{3/2} - m_{Z^\prime}(T)^{3/2}\right).
    \label{eq:Veff}
\end{split}\end{equation}
The second line corresponds to the Coleman-Weinberg potential plus counter-terms added to ensure that the minimum of the 1-loop zero-temperature potential remains at $v_s$, and that the mass of the scalar is still given by~\eq{eq:ms}. We take into account the $Z^\prime$ running in the loop, and neglect the right-handed neutrinos (assuming small Yukawas\footnote{Including the right-handed neutrinos would amount to shifting the prefactor in the second line of \eq{eq:Veff} to $3\!\times\! (2g_{BL})^4 - 2\!\times\! \sum_i (y_i^M)^4/4$. Here we assume $\sum_i (y_i^M)^4 \ll 96 g_{BL}^4$.}) and the scalar (since typically $\lambda_s \ll g_{BL}$ and it has only one degree of freedom, so its effect is subdominant against the $Z^\prime$). The thermal part $V_\text{th}$ is computed at 1-loop via standard methods of thermal field theory~\cite{Dolan:1973qd, Hindmarsh:2020hop}, and the last term accounts for the resummation of daisy diagrams, with the thermal mass for the $Z^\prime$ given by
\begin{equation}
    m^2_{Z^\prime}(T) = 4g^2 \phi_s^2 + \dfrac{17}{6}g^2 T^2.
\end{equation}

Notice that the value of the zero-temperature potential at the unbroken state $\phi_s=0$ is parametrized by $\lambda_s$, whereas at the broken vacuum $\phi_s=v_s$ it goes with $g_{BL}$. Hence, for too small values of $\lambda_s$ the symmetric minimum may become the lowest energy state and $U(1)_{B-L}$ would remain unbroken, which is nonphysical and should be avoided. Explicitly, defining the energy difference between the two vacua $\Delta V(T,\phi)\equiv V_\text{eff}(T,\phi)-V_\text{eff}(T,0)$, one finds that at zero-temperature
\begin{equation}
    \Delta V(0,v_s) = \dfrac{3}{8\pi^2} g_{BL}^4 v_s^4 - \dfrac{\lambda_s}{8}v_s^4 < 0,
    \label{eq:dV}
\end{equation}
implying $\lambda_s > 3g_{BL}^4/\pi^2$.

The bubble nucleation rate per unit volume is $  \Gamma_\text{nuc}/\mathcal{V} \sim T^4 e^{-S_3/T}$, 
where 
\begin{equation}
    S_3 = 4\pi\int dr\,r^2\left[\frac{1}{2}\left(\dfrac{d\phi_c}{dr}\right)^2 + \Delta V(T, \phi_c)\right]
\end{equation}
is the Euclidean action of the critical bubble, corresponding to the saddle point configuration that connects the two vacua in field space~\cite{Hindmarsh:2020hop, Mukhanov:2005sc}. At some temperature $T$, 
 this configuration can be found by solving the bounce equation $-\nabla^2 \phi + dV_\text{eff}/d\phi=0$ using a shooting method. The nucleation temperature is found by imposing that $\Gamma_\text{nuc}/\mathcal{V}$ equals the Hubble expansion rate. This corresponds roughly to the temperature at which $S_3/T\approx 140$~\cite{Caprini:2019egz}. We can then define the fractional amount of energy released by the transition~\cite{Espinosa:2010hh} to be,
\begin{equation}
    \alpha \equiv \dfrac{Q_\text{lat}-3\Delta V}{4\rho_\text{rad}},
\end{equation}
where $Q_\text{lat}=Td\Delta V/dT - \Delta V$ is the latent heat of the phase transition, the numerator $Q_\text{lat}-3\Delta V$ is the difference of the trace of the energy-momentum tensor in both phases~\cite{Espinosa:2010hh, Giese:2020rtr}, $\rho_\text{rad}=\pi^2 g_\text{eff}T^4/30$ is the radiation energy density at the time of the transition and $g_\text{eff}=117.5$ is the effective number of relativistic degrees of freedom in the plasma. The amplitude of the GW spectrum will depend on how efficiently this energy is converted into the fluid motion of the plasma. For that, we construct efficiency factors $\kappa_\text{sw}$ and $\kappa_\text{turb}$ such that the fractional energy in sound waves (respectively in plasmatic turbulence) is proportional to $\kappa_\text{sw}\alpha$ (resp. $\kappa_\text{turb}\alpha$). An approximate formula for $\kappa_\text{sw}$ can be found in ref.~\cite{Espinosa:2010hh}, but for turbulence this conversion factor is unknown. Sometimes it is estimated to be $1-10\%$ of $\kappa_\text{sw}$~\cite{Caprini:2019egz, Caprini:2015zlo}, and we have explicitly checked that for these values its contribution is subdominant. Due to this indeterminacy and subdominance, we neglect here the contribution from turbulence and consider only sound waves. Moreover, we take into account the suppression factor of the sound wave spectrum due to the finite lifetime of this source and a possibly early onset of turbulence~\cite{Guo:2020grp}. Altogether, this means that our approach {underestimates} the spectrum, so our results are conservative.

Another important parameter for estimating the GW spectrum is the (inverse) duration of the transition, estimated as\footnote{For the phase transitions considered here, $\beta/H \sim 100-1000 \gg 1$. The spectrum of GWs from sound waves scales as $(\beta/H)^{-1}$ whereas bubble collisions is damped by $(\beta/H)^{-2}$, which is comparatively negligible~\cite{Caprini:2015zlo}.}
\begin{equation}
    \dfrac{\beta}{H}\equiv T\dfrac{d(S_3/T)}{dT}.
\end{equation}
Given that the bubble expands at a velocity $v_w$, its radius at collision will be proportional to $v_w (H/\beta)$, and the larger the bubble, the larger the GW amplitude. Calculating the wall velocity $v_w$ is a daunting task since it involves non-equilibrium phenomena and depends on how we model the plasma away from equilibrium. There have been recent discussions in the literature on the appropriate way to achieve this description, but the debate is still unsettled~\cite{Cline:2020jre, Dorsch:2021nje, Laurent:2022jrs}. Here we approximate $v_w=1$ for simplicity, which should suffice for an adequate estimate of the spectra at the correct order of magnitude.

More specifically, for the amplitude of the GW spectrum from sound waves we find~\cite{Caprini:2015zlo, Hindmarsh:2015qta, Hindmarsh:2017gnf},
\begin{equation}\begin{split}
    \Omega_\text{sw} h^2 = &~2.65\times 10^{-6}\,v_w\left(\dfrac{H}{\beta}\right)\left(\dfrac{\kappa_\text{sw}\alpha}{1+\alpha}\right)^2 \left(\dfrac{100}{g_\text{eff}}\right)^{1/3} \mathcal{Y}\times\\
    &\quad\times \left(\dfrac{f}{f_\text{peak}}\right)^3\left(\dfrac{7}{4+3(f/f_\text{peak})}\right)^{7/2},
\end{split}\end{equation}
where 
\begin{equation}
    \mathcal{Y} = 1- ( 1 + 2\tau_{\text{sw}} H)
\end{equation}
is the suppression factor due to the finite lifetime $\tau_\text{sw} = R/\overline{U}$ of the sound wave source, where $R = (8\pi)^{1/3}v_w/\beta$ is the typical bubble separation and  $\overline{U} = \left(\frac34\frac{\kappa_\text{sw}\alpha}{1+\alpha}\right)^{1/2}$ the plasma root mean squared velocity~\cite{Guo:2020grp, Hindmarsh:2015qta, Hindmarsh:2017gnf}.

This spectrum has a peak at
\begin{equation}
    f_\text{peak} = 1.9\times 10^{-2}\text{mHz}\,\left(\dfrac{1}{v_w}\dfrac{\beta}{H}\right)\left(\dfrac{T}{100~\text{GeV}}\right)\left(\dfrac{g_\text{eff}}{100}\right)^{1/6}.
\end{equation}

We will see that, for typical benchmark values of the parameters in our model, the peak frequency lies in the mHz band, hence possibly within reach of future interferometers such as LISA, DECIGO and BBO. We will now discuss the dark matter relic density and scattering rate to later put our findings into perspective.

\section{\label{sec:dmproduction}Dark Matter Relic Abundance}
In this section, we assess the production of DM particles in the standard thermal freeze-out paradigm. In such a narrative, after reheating, the DM particles were in thermal equilibrium with the SM particles, which means that they were pair-annihilated and pair-produced in equal rate in the early universe. However, the universe is expanding and cooling down. Although both the Hubble rate and the interaction rate are decreasing, at a given temperature, the Hubble rate overcomes the interaction rate, and freeze-out takes place fixing the DM relic abundance.

In this paradigm, the evolution of the DM number density $n_{DM}$ is described by the Boltzmann equation, 
\begin{equation}\label{eq:BEQ}
   \frac{dY_{DM}(x)}{dx} =  - \frac{s(x) \langle \sigma v \rangle }{x\,H(x)}\left[Y^2_{DM}(x) - Y^{\textrm{eq}\,2}_{DM}(x)\right]
\end{equation}
where $Y_{DM} = n_{DM} /s$ is the comoving number density, with 
\begin{equation}
    s(x) = \frac{2\pi^2}{45}g_{\star s}(x)m^3_{DM} x^{-3}
\end{equation}
representing the entropy of the primordial plasma, with $g_{\star s}$ being the relativistic degrees of freedom that contribute to the entropy, and $x = m_{DM}/T$ is a ``time'' variable that helps us to simplify the integration and physical interpretations. In this parametrization, the Hubble expansion rate and comoving abundance are written as, 
\begin{equation}
    H(x) = \frac{\pi}{M_{pl}}\sqrt{\frac{g_\star(x)}{90}}m_{DM}^2 x^{-2} ,
\end{equation}

\begin{equation}
    Y^{\textrm{eq}}_{DM}(x) = \frac{45}{4\pi^4}\frac{g_{DM}}{g_{\star s}}x^2 K_2 (x),
\end{equation} where $g_\star$ accounts for the relativistic degrees of freedom,
 with $g_{\star s}  \approx g_\star= 106.75$ at the time of freeze-out, $g_{DM}$ corresponds to the DM degrees of freedom, and $K_2(x)$ is the modified Bessel function. After solving Eq.~\eqref{eq:BEQ}, we obtain the DM abundance,
\begin{equation}
\Omega_{DM} h^2 \simeq 2.82\times10^8 m_{DM} Y_{DM}(x \rightarrow \infty),
\end{equation}
where we use the limit of $x \rightarrow \infty$ in the solution of the Boltzmann equation. The most current value is given by Planck collaboration, $\Omega_{DM} h^2 =  0.1200 \pm 0.0012$ within 68\% C.L. \cite{Planck:2018vyg}. 

\begin{figure}
    \centering
    \includegraphics[width=1\linewidth]{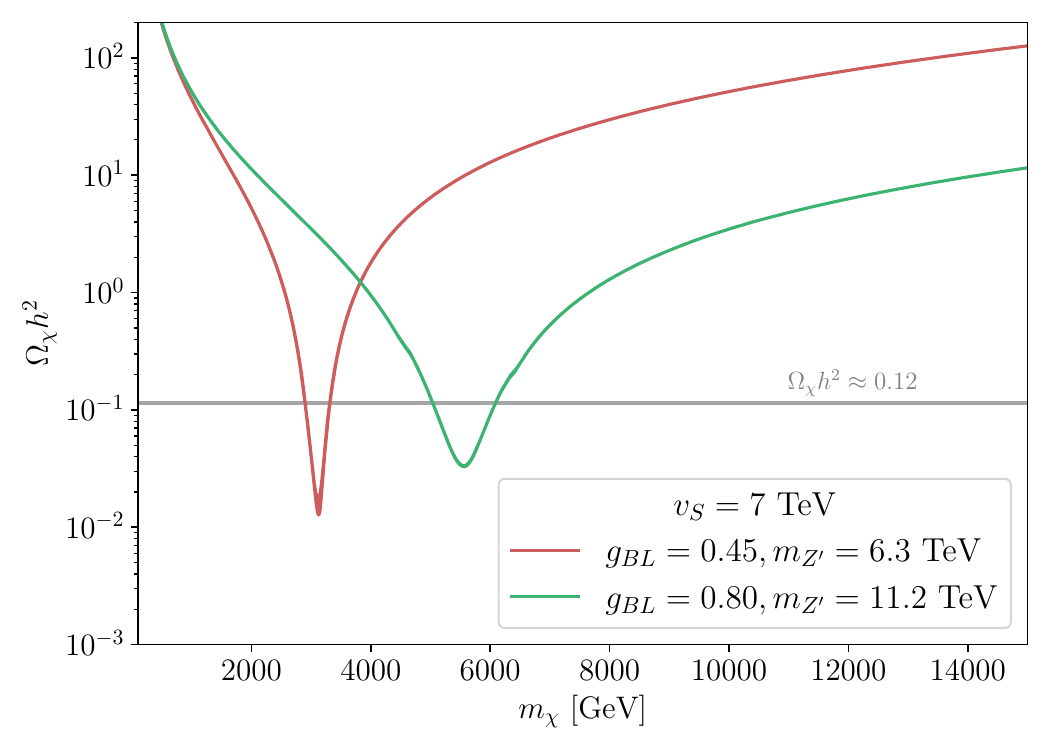}
    \caption{DM relic abundance as a function of the DM mass. We set $v_s = 7$~TeV. The red solid curve is the relic abundance for $g_{BL} = 0.45$ and $m_{Z^\prime} = 6.3$~TeV, while the green one represents $g_{BL} = 0.80$ and $m_{Z^\prime} = 11.2$~TeV. The gray line is the abundance in agreement with Planck observations \cite{Planck:2018vyg}. Notice that there are two viable DM masses.}
    \label{fig:omegaxmdm}
\end{figure}

In Section~\ref{sec:model}, we described how a vector-like Dirac fermion $\chi$ can be the DM candidate in a minimal $B-L$ model. We computed the DM relic density using micrOMEGAs \cite{Belanger:2006is, Belanger:2020gnr}. In Fig.~\ref{fig:omegaxmdm}, we present the behavior of the relic abundance as a function of the DM mass. We have set $v_s =7$~TeV. The red and green solid curves are the relic abundance for $g_{BL} = 0.45$ and $g_{BL} = 0.80$ (corresponding respectively to $m_{Z^\prime} = 6.3$~TeV and $m_{Z^\prime} = 11.2$~TeV, according to Eq.~\eqref{eq:zpmasseq}). The gray horizontal line delimits the region that reproduces the Planck data \cite{Planck:2018vyg}. 

It is remarkable that the largest part of the parameter space of the thermal relic abundance is overabundant. It occurs because, in general, the model provides small annihilation cross-sections. However, at the resonance regime, when $m\chi \approx m_{Z^\prime}/2$, it reaches sizeable values via the s-channel in Fig.~\ref{fig:feynmandigrams} (a). Such an enhancement brings the relic density down to the observed value. Because of the (inverted) resonance peak, we see that there are typically two viable DM masses: one slightly smaller than $m_{Z^\prime}/2$, the other slightly larger.

Thus far, we have addressed how to produce dark matter and GWs. In what follows, we will show that our model is amenable to collider and direct detection constraints.

\section{\label{sec:bounds} The Constraints}

Let us now turn to a discussion of the limits on the $Z^\prime$ mass from the ATLAS and LEP-II results, and direct detection bounds on the spin independent (SI) cross-section as reported by XENONnT and LUX-ZEPLIN (LZ).

\subsection{\label{sec:collider} Collider limits}

Extra gauge bosons with unsuppressed coupling to the SM fermions, as the one considered in this work, produce strong collider signals in the form of resonances decaying into SM pairs as for example dileptons \cite{ATLAS:2019erb}, dijets \cite{ATLAS:2019fgd} and di-top \cite{ATLAS:2019npw}. We will adopt the most restrictive limits to our model, which stem from searches for heavy dilepton events conducted by ATLAS collaboration during Run 2 of the Large Hadron Collider, and corresponds to an integrated luminosity of $139fb^{-1}$ \cite{ATLAS:2019erb}. It is well-known that the presence of $Z^\prime$ couplings to DM pairs can weaken the latter constraints \cite{Arcadi:2013qia}, however, such invisible decay yields no effect on the dilepton bound, because of its small contribution to the total $Z^\prime$ width \cite{Klasen:2016qux}. ATLAS collaboration has not searched for a B-L $Z^\prime$ boson. Therefore, we had to derive our own limit by comparing the theoretical prediction with the upper limit derived at 95\% CL on the fiducial cross-section times branching ratio quoted in \cite{ATLAS:2019erb}. 

For each combination of gauge coupling and $Z^\prime$ mass, we computed the $Z^\prime$ branching ratio using CalcHEP \cite{Belyaev:2012qa} and fed this information into Madgraph5 \cite{Alwall:2014hca,Frederix:2018nkq} where we performed the Monte Carlo simulation adopting a parton distribution function (PDF) NNPDF23LO \cite{Carrazza:2013axa}. We computed the signal event $pp \to Z^{\prime} \to \ell\bar{\ell}$ at  $\sqrt{s}=13$ TeV, with $\ell = e,\mu$ and compared our result with the public result from ATLAS Collaboration \cite{ATLAS:2019erb}. Following the collaboration, we required the signal events to feature two opposite charge leptons, with transverse momentum $p_{\mathrm{T}}>30$~GeV, and pseudorapidity $|\eta|<2.5$. In doing so, we obtained the collider limits shown in Table~.\ref{tablecollider}. We emphasize that these findings represent a new result in the literature. We highlight that the limit for $g_{BL}=0.8$, the production cross-section times branching ratio falls outside the sensitivity of ATLAS data. This happens because when we increase the gauge coupling, we also increase the production cross section of $Z^\prime$ bosons. This shift upward in the cross-section makes the theoretical prediction cross the exclusion limit from ATLAS at larger $Z^\prime$ masses. As the highest pole mass probed by ATLAS was $6$~TeV, when we significantly increase the gauge coupling, our model can no longer be excluded by it. For this particular case, we applied the Collider Reach $\beta$ tool \cite{Thamm:2015zwa} which allows us to forecast the new bounds on the $Z^\prime$ mass for a different collider configuration. Here we maintained $\sqrt{s}=13$~TeV, and simply ramped up the luminosity and selected to be $\mathcal{L}=300 fb^{-1}$. Doing so, we project the ATLAS bound of $m_{Z^\prime}> 7.7$~TeV. This is a simple estimation of the ATLAS reach. For $g_{BL} =0.8$, we could have adopted a more solid and weaker bound, $m_{Z^\prime}> 5.6$~TeV, from the Large Electron Positron (LEP) collider that will address below.

\begin{table}[]
    \centering
    \begin{tabular}{|c|c|}
     \hline
    Gauge coupling &  Lower bound - ATLAS 13TeV\\
    \hline
        $g_{BL}=0.2$ &  $m_{Z^\prime}> 4.94$~TeV\\
        $g_{BL}=0.3$ &  $m_{Z^\prime}> 5.35$~TeV\\
        $g_{BL}=0.4$ &  $m_{Z^\prime}> 5.62$~TeV\\
        $g_{BL}=0.45$ &  $m_{Z^\prime}> 5.75$~TeV\\
        $g_{BL}=0.5$ &  $m_{Z^\prime}> 5.8$~TeV\\
        $g_{BL}=0.6$ &  $m_{Z^\prime}> 5.97$~TeV\\
        $g_{BL}=0.7$ &  $m_{Z^\prime}> 6$~TeV\\
        $g_{BL}=0.8$ &  $m_{Z^\prime}> 7.7$~TeV\\
    \hline
    \end{tabular}
    \caption{Lower mass bounds derived on the $Z^\prime$ mass for different gauge couplings using ATLAS results reported in \cite{ATLAS:2019erb}.}
    \label{tablecollider}
\end{table}

An old and relevant collider bound comes from LEP-II data that reads $m_{Z'}/g_{BL}> 7\,\mbox{TeV}$ \cite{Carena:2004xs,Cacciapaglia:2006pk}. This limit is not affected by an eventual presence of an invisible branching fraction. It stems from the comparison between the SM prediction and new physics prediction for dilepton events, rather than from resonance searches. As LEP featured fantastic precision due to the leptonic nature of the process, they were able to obtain a stringent limit on the $Z^\prime$ mass. LEP bound is more relevant for $g_{BL}\sim 1$. We explore a setup with  $g_{BL}=0.8$, and as we discussed previously, current data from ATLAS cannot probe this case. For this reason, for $g_{BL}=0.8$, we use the limit from LEP which reads $5.6$~TeV.

\subsection{\label{sec:dd} Direct detection bounds}

Since the $Z^{\prime}$ boson features vector coupling with both quark and DM pairs, spin-independent interactions arise between the latter and the nucleons. The corresponding cross-section can be written as
\begin{equation}
   \sigma_{\chi N=p,n}^{\rm SI} = \frac{\mu_{\chi N}^2}{\pi}\frac{9 \, n_q^2 n_\chi^2 g_{BL}^4}{m_{Z^{\prime}}^4}\label{eq:csSI}
\end{equation}
with $\mu_{\chi N}=\frac{m_\chi m_N}{m_\chi+m_N}$ being the DM-nucleon reduced mass.
Such interactions are strongly constrained by xenon-based direct detection experiments. For our study, we will consider the most recent bounds as given by the LZ \cite{LZ:2022ufs} and XENONnT \cite{XENON:2023sxq}. Notice also that the vector coupling with SM fermions implies an s-wave dominated annihilation cross-section into SM fermions, hence potentially testable via indirect detection. One could then use searches of $\gamma$-ray signals, see e.g. \cite{Fermi-LAT:2015att,CTA:2020qlo,Hooper:2012sr,CTAConsortium:2017dvg}, to further constrain the parameter of the model. Indirect detection constraints are, however, not competitive with direct detection and collider for the model under scrutiny. Consequently, we will not show them explicitly. That said, we will now put our findings into perspective.

\section{\label{sec:results} Results}

\begin{figure}[h!]
    \includegraphics[width=.9\linewidth]{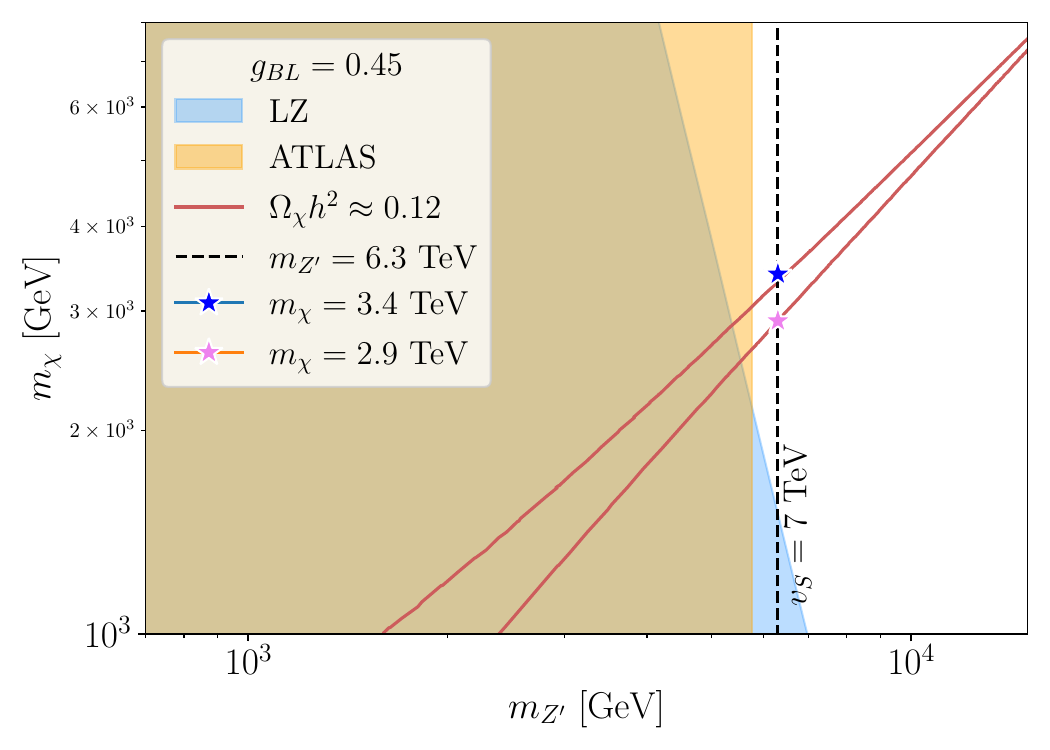}\\
    \includegraphics[width=.9\linewidth]{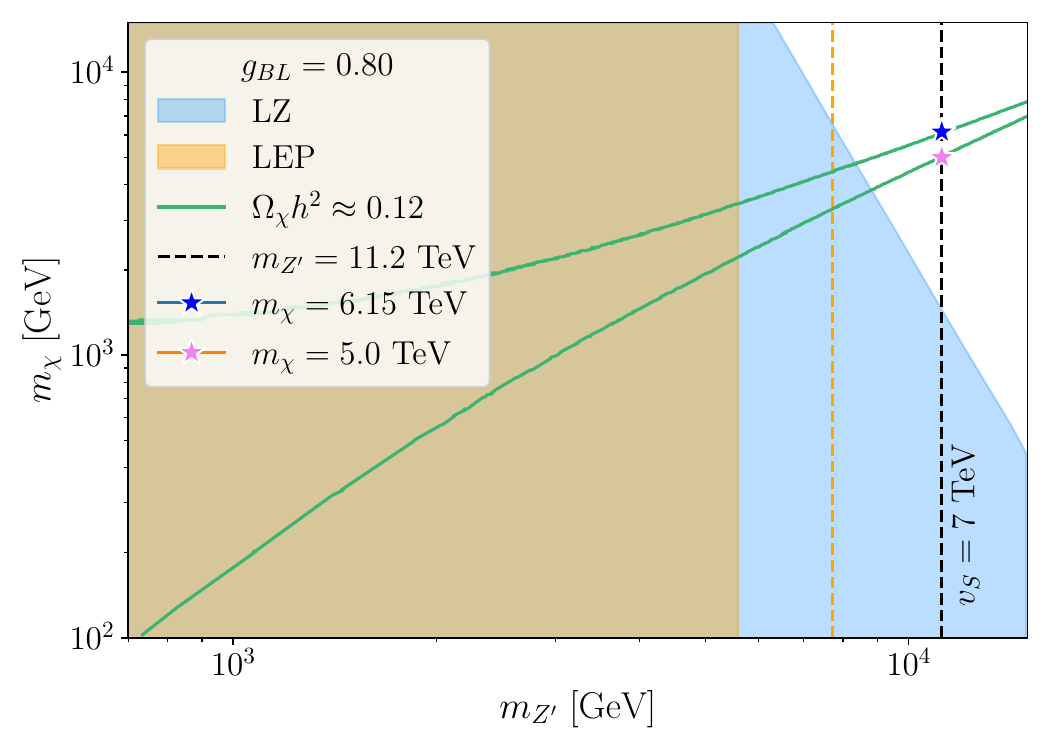}
    \caption{Combined results for DM phenomenology. The solid curves correspond to the correct DM relic abundance $\Omega_\chi h^2 \approx 0.12$, for $g_{BL} = 0.45$ (top) and $g_{BL} = 0.80$ (bottom). The region between the two curves corresponds to underabundance and is in principle allowed, while outside the curves the Universe would overclose. The shaded blue region represents the parameter space excluded by direct detection. In orange, we have the corresponding collider limits, which substantially constrain the model.}
    \label{fig:abundance}
\end{figure}

Our main results concerning the DM phenomenology are displayed in Fig. \ref{fig:abundance} for $g_{BL} = 0.45$ (top panel) and $g_{BL} = 0.80$ (bottom panel). The red (top) and green (bottom) curves yield the correct relic abundance, in agreement with Planck's observations. The region in between the curves gives an underabundant relic, whereas the outside region is an overabundant one. Furthermore, notice that one value of $Z^\prime$ mass is associated with two values of DM mass that yield the correct relic density. This happens because the model reaches $\Omega_{\chi}h^2 \approx 0.11$ in the $Z^\prime$ resonance regime.

We also exhibit the direct detection constraint from the LZ collaboration that is slightly stronger than the XENONnT one. The shape of the experiment curve can be understood looking at Eq.~\eqref{eq:csSI}. When the dark matter mass is much larger than the nucleus mass, the scattering cross-section is independent of the dark matter mass, but decreases with $m_{Z^{\prime}}^4$. As the experimental limit linearly weakens with the dark matter, the viable parameter space in the $m_{\chi}$ vs $ m_{Z^\prime}$ plane is mostly sensitive to the $Z^\prime$ mass. We are plotting our findings in a Log-Log scale, thus what we see is a line representing the direct detection bound weakening as the $Z^\prime$ mass increases. 

As for the collider constraint from ATLAS, $m_{Z^\prime} > 6$~TeV. It is simply a vertical line on the $Z^\prime$ mass once we fix the gauge coupling. This has to do with the fact that the signal scales with $g_{BL}^2\times BR(Z^\prime \rightarrow l\bar{l})$. Hence, once we fix $g_{BL}$ the bound on the $Z^\prime$ can be directly extracted from the collaboration report. The branching of the $Z^\prime$ into dileptons will not change when $m_{Z^\prime} > 2 m_{\chi}$, because the partial width of $Z^\prime \rightarrow \chi \bar{\chi}$ is also controlled by $g_{BL}$. Therefore, when we open the invisible decay into dark matter, no meaningful change in the branching into lepton is expected. In models, where the $Z^\prime$ coupling to dark matter is different from the $Z^\prime$ coupling to leptons, a large decay into invisible can be more pronounced, however, \cite{Alves:2013tqa,Alves:2015pea,Alves:2015mua}.

\begin{figure}[h!]
    \includegraphics[width=.9\linewidth]{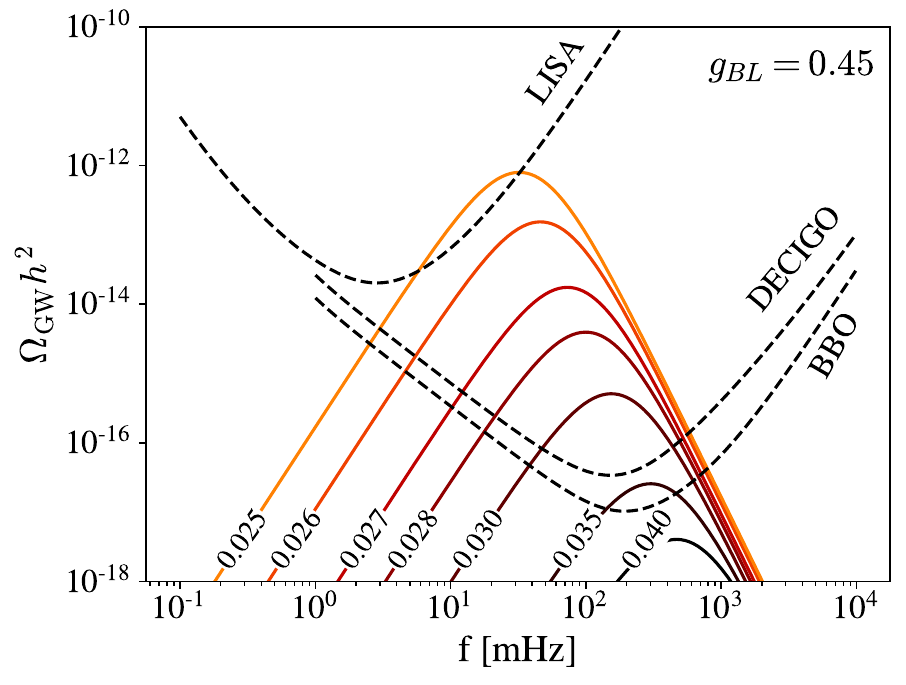}\\
    \includegraphics[width=.9\linewidth]{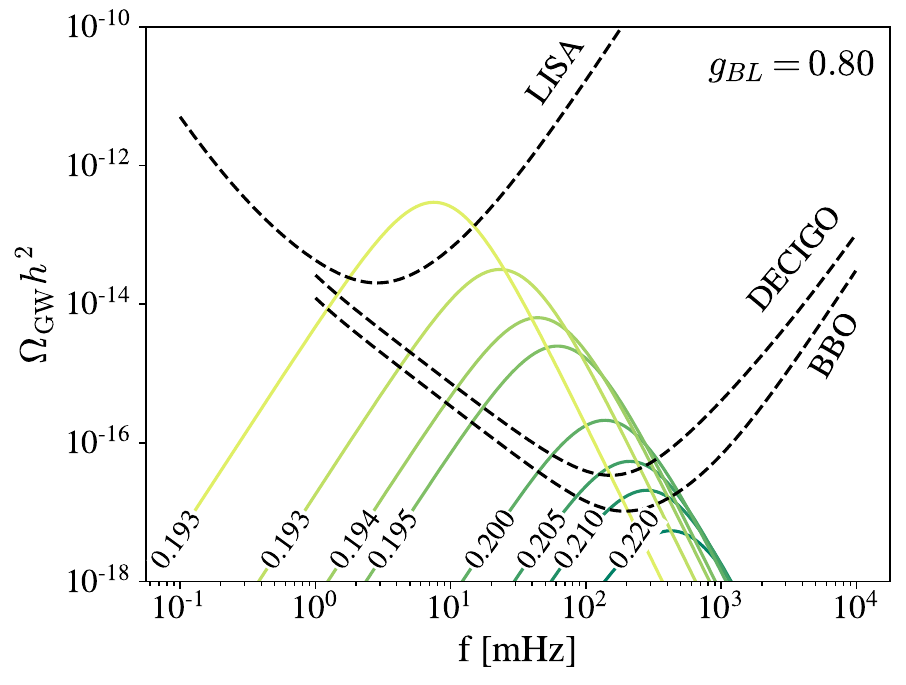}
    \caption{Spectra of GWs for $v_s=7$~TeV. The figure at the top (respectively bottom) refers to $g_{BL}=0.45$ (resp. $g_{BL}=0.80$). The various spectra displayed are obtained by varying $\lambda_s$ at values shown in each curve. Sensitivity curves for LISA, DECIGO and BBO have been obtained from Ref.~\cite{Schmitz:2020syl}.}
    \label{fig:spectra}
\end{figure}

Notice that in Fig.~\ref{fig:abundance},  $v_s$ varies because the $Z^\prime$ mass changes linearly with it. We are bringing this to the reader's attention because the value of $v_s$ will be rather relevant to the GW spectra. For this reason, we drew a dashed vertical line representing $v_s = 7$~TeV, which realizes the GW spectra to be explored later in  Fig.~\ref{fig:spectra}. In the same way, we benchmark with blue and violet stars DM masses associated to those GW spectra. In the top panel, $m_\chi = 2.9$ or $3.4$~TeV, while in the bottom one $m_\chi$ has to be $5.0$ or $6.15$~TeV. The shift upward in the relic density curve has to do with the $Z^\prime$ width, which is larger as it grows with $g_{BL}$.

\begin{figure}
    \includegraphics[width=.9\linewidth]{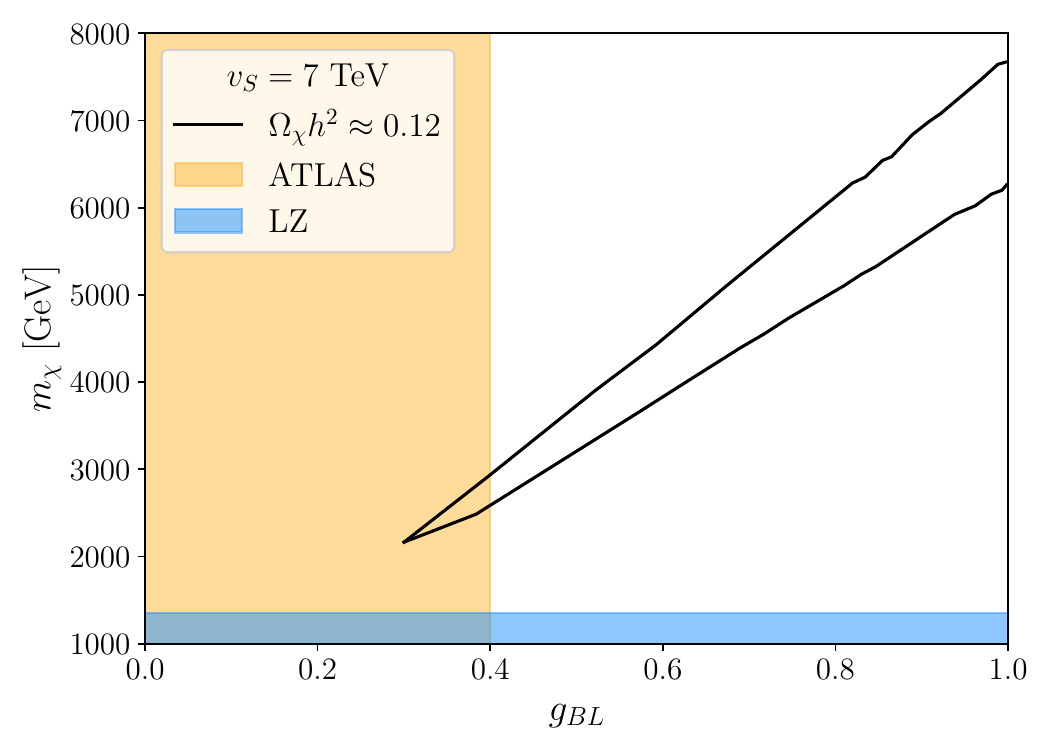}
    \caption{The observed DM relic for $v_S = 7$~TeV is represented by the black curve. The shaded orange and blue regions are the ATLAS and LUX-ZEPLIN limits, respectively. The ATLAS bound reproduces Table~\ref{tablecollider}. Direct detection imposes $m_\chi > 1.35$~TeV.}
    \label{fig:mchixgbl}
\end{figure}

In Fig.~\ref{fig:mchixgbl} we fix $v_s = 7$~TeV and obtain the relic density curve in the $ m_\chi$ vs $g_{BL}$ plane, whereas $m_{Z'}$ varies. The shaded region is ruled out by the LZ collaboration. From Eq.~\eqref{eq:csSI} we notice that the DM-nucleon SI scattering cross-section will depend only on $m_
\chi$, since the dependence on $g_{BL}$ enters only in the ratio $\left(g_{BL}/m_{Z^\prime}\right)^4$, which is constant for fixed $v_s$. As we are considering $m_{\chi} \gg m_N$, the scattering cross-section no longer depends on the dark matter mass, and is consequently constant in the plane of Fig.~\ref{fig:mchixgbl}. Although, the experimental limit from direct detection experiments linearly weakens with the dark matter mass. In particular, we found a dark matter-nucleon scattering cross section of $3.1 \times 10^{-10}$~pb, which is consistent with the experiment limit that reads $3.2 \times 10^{-10}$~pb for $m_{\chi} \simeq 1.35$~TeV. Therefore, for larger dark matter masses, the direct detection bound becomes too weak to constrain the parameter space in Fig.~\ref{fig:mchixgbl}. This explains why the exclusion region from direct detection represents a horizontal blue curve. Collider bounds are shown following Table~\ref{tablecollider}.

The above discussion highlights how DM searches can shed light on the properties of the DM particle $\chi$, the $Z^\prime$ mediator, and the gauge coupling $g_{BL}$. But these searches are basically insensitive to
the properties of the scalar $\phi_s$, such as its mass or self-coupling $\lambda_{s}$, which do not affect the DM density. Moreover, if $m_{\phi_s} \lesssim 2 m_{Z^\prime}$ (i.e. $\lambda_s \lesssim 16 g_{BL}^2$), the scalar will only decay to the invisible right-handed neutrinos and will be hardly detectable at colliders. Fortunately, in this case, one might constrain the scalar sector using gravitational waves, and an eventual detection of GWs would allow us to determine the scalar self-coupling.

This is shown in \fig{fig:spectra}, where we display the GW spectra for the benchmark value of  $v_s=7$~TeV with $g_{BL}=0.45$ (top) and $g_{BL}=0.80$ (bottom). The different spectra in each figure correspond to varying $\lambda_s$, whose corresponding values are shown along each curve. Also shown are sensitivity curves for LISA, DECIGO and BBO~\cite{Schmitz:2020syl}. They have been constructed by taking into account the expected shape of the signal from a first order first transition, such that if the predicted spectrum lies anywhere above the curve, the expected signal-to-noise ratio is larger than 1. Notice that decreasing $\lambda_s$ generates larger spectra. This is because smaller $\lambda_s$ will lead to a smaller energy difference in the zero-temperature potential between the broken and unbroken minima (cf.~\eq{eq:dV}), which leads to stronger phase transitions~\cite{Dorsch:2017nza}. Eventually, the energy gap is so small that the transition to the true minimum never occurs: the field remains trapped in the metastable false vacuum and the symmetry is not broken, which is obviously non-physical. Hence, there is an upper bound on the spectrum that could be achieved for each pair $(v_s, g_{BL})$.

Varying $v_s$ will shift the peak frequency, but since the plot is logarithmic in $f$, one would need large differences in $v_s$ for this shift to be noticeable. The differences in the spectra will not be significant for the range of $v_s$ values considered here (cf. the allowed region of DM production in \fig{fig:abundance}). 

\begin{figure}[h!]
    \includegraphics[width=.9\linewidth]{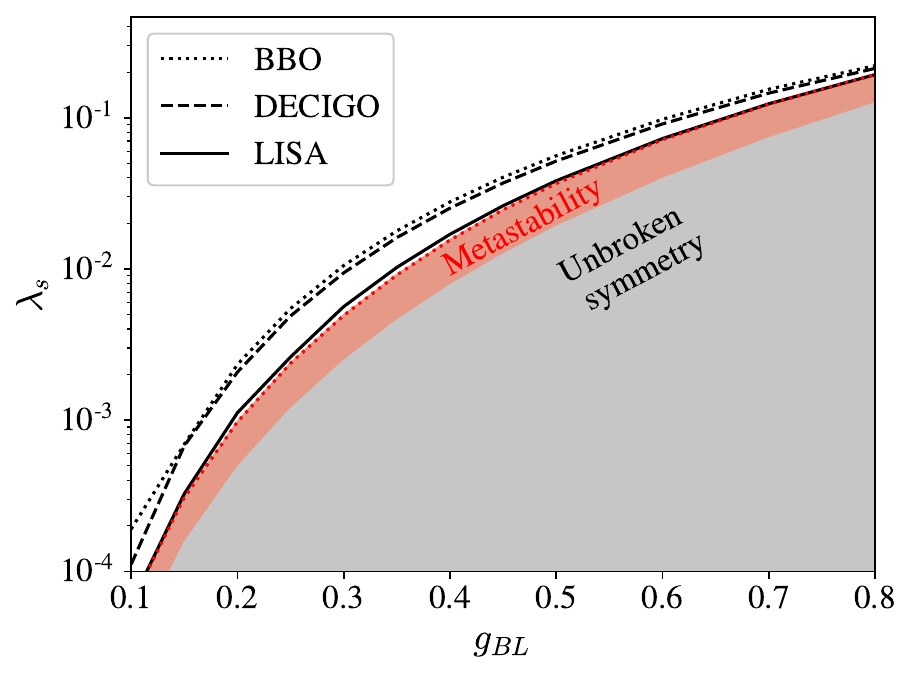}
    \caption{Maximum values of $\lambda_s$, for given $g_{BL}$ and $v_s=7$~TeV, that would yield a GW spectrum detectable at BBO (dotted curve), DECIGO (dashed) and LISA (solid). The red region shows the values of $\lambda_s$ that lead to a metastable false vacuum, whereas the gray region corresponds to stability of the unbroken state, i.e. the broken phase not being the global minimum of the potential, as per \eq{eq:dV}.}
    \label{fig:lmax}
\end{figure}

\fig{fig:lmax} shows the maximum values of $\lambda_s$, for given $g_{BL}$ and fixed $v_s=7$~TeV, that would yield a detectable spectrum at BBO, DECIGO and LISA, respectively. This is computed by finding the value of $\lambda_s$ at which the predicted spectrum intersects the sensitivity curve at any point. In other words, the parameter space below the curves could be probed using gravitational waves. We also show the regions where the unbroken state is metastable (red region) and stable (gray region). For concreteness, considering BBO sensitivity, the region that lies between the dotted curve and the metastability yields a detectable gravitational wave signal. Thus, the detectability at LISA is achieved very close to the metastable situation. This means that, for the perturbative range of $g_{BL}$ shown in the plot, one can find spectra within the LISA sensitivity, but not by a huge margin. More sensitive detectors, such as DECIGO and BBO, would be able to probe a larger range of $\lambda_s$ values.

\section{Discussions}

Figure~\ref{fig:spectra} shows that the GW spectra are highly sensitive to $\lambda_s$ and to $g_{BL}$. There is also a dependence on $v_s$ (and hence on $m_{Z^\prime}$), which is not shown explicitly in the figure, but can be understood from the fact that this parameter sets the characteristic energy scale of the transition and therefore determines the characteristic frequency of the spectral peak. Hence, from a measurement of the GW spectrum alone, one cannot determine any of these parameters separately. On the other hand, $g_{BL}$, $m_{Z^\prime}$ and $m_\chi$ govern the dark matter phenomenology, and could be probed by colliders and direct detection experiments, as shown in figures~\ref{fig:abundance} and \ref{fig:mchixgbl}. Therefore, one really needs to use colliders, direct detection \textit{and} GW experiments to gain access to multiple directions of the parameter space independently.

It should be mentioned that the GW spectra shown in Fig.~\ref{fig:spectra} are subject to theoretical uncertainties due to the 1-loop approximation used for the effective potential. The various sources of uncertainties have been studied in ref.~\cite{Croon:2020cgk} in the context of a $\phi^6$ effective theory, where it was shown that the theoretical prediction for the GW amplitude at 1-loop could vary up to 2 orders of magnitude for that model. Taking this figure at face value, disregarding the striking differences between the models, we conclude from Fig.~\ref{fig:spectra} that this would imply an uncertainty of $\mathcal{O}(10\%)$ in our mapping from self-couplings to GW spectra. In this paper, we do not aim at such level of precision. However, once such GWs are detected, theoretical accuracy will definitely be desired. For further discussion on this matter, see also~\cite{Ekstedt:2022bff, Schicho:2022wty, Lofgren:2021ogg}.

\section{\label{sec:conclusions} Conclusions}
In this work, we have argued that GW detectors have to be allied to collider and direct detection experiments in order to be able to probe the full parameter space of models with a dark sector. Dark scalars are typically extremely difficult to detect at colliders, and measuring their self-couplings are unimaginable with the machines we will build in the near-future. However, the GW spectrum from a first order phase transition is highly sensitive to this scalar sector, and depend on the scalar self-coupling as well as on the gauge and fermionic sectors. Our main point is that, in order to disentangle the parameter dependence on these observables, one must consider collider, direct detection and gravitational wave experiments. 

We illustrate this point concretely in a minimal $B-L$ model with a viable DM candidate. The DM candidate is a vector-like fermion and the $U(1)_{B-L}$ symmetry is broken by a scalar singlet and induces a first-order cosmological phase transition. Interestingly, we find that the expected LISA sensitivity is barely enough to probe the most extreme cases, when the phase transition is extremely strong and the unbroken vacuum is on the verge of becoming metastable. On the other hand, more sensitive interferometers such as DECIGO and BBO might be able to probe a larger range of self-couplings. There is arguably some degree of tuning for gravitational wave detection, in the sense that an exiguous change in the value of $\lambda_s$ might move the spectrum from within the detectability range of LISA down to non-detectability even at BBO. But this also means that a gravitational wave detection, allied with measurements of $v_s$ and $g_{BL}$ from other experiments, would lead to a measurement of the scalar self-coupling with good precision.

Colliders can probe the $Z^\prime$ mass and the gauge coupling $g_{BL}$, but are insensitive to details of the scalar singlet. Direct detection experiments rule out a larger fraction of the parameter space, which helps us constrain the dark sector. Knowing that, if we impose a gravitational wave detection in future probes and thermal production of dark matter, we can predict which dark matter masses reproduce the correct relic density in agreement with the data. In summary, our main result is that gravitational wave detectors offer a complementary and orthogonal probe to dark sectors, allowing us to further narrow down the parameter space of dark matter models. 

\acknowledgments
The authors thank Diego Cogollo, Carlos Pires, Carlos Yaguna for discussions. JPN acknowledges support from Coordenaç\~ao de Aperfeiçoamento de Pessoal de N\'ivel Superior (CAPES) under Grant No. 88887.712383/2022-00. Y.M.O.T.
acknowledges financial support from CAPES under grants
88887.485509/2020-00. This work was financially supported by Simons Foundation (Award Number:1023171-RC), FAPESP Grant 2021/01089-1, ICTP-SAIFR FAPESP Grants 2021/14335-0, CNPq Grant 307130/2021-5, FONDECYT Grant 1191103 (Chile) and ANID-Programa Milenio-code ICN2019\_044.

\nocite{*}
\bibliography{ref}

\begin{thebibliography}{68}%
\makeatletter
\providecommand \@ifxundefined [1]{%
 \@ifx{#1\undefined}
}%
\providecommand \@ifnum [1]{%
 \ifnum #1\expandafter \@firstoftwo
 \else \expandafter \@secondoftwo
 \fi
}%
\providecommand \@ifx [1]{%
 \ifx #1\expandafter \@firstoftwo
 \else \expandafter \@secondoftwo
 \fi
}%
\providecommand \natexlab [1]{#1}%
\providecommand \enquote  [1]{``#1''}%
\providecommand \bibnamefont  [1]{#1}%
\providecommand \bibfnamefont [1]{#1}%
\providecommand \citenamefont [1]{#1}%
\providecommand \href@noop [0]{\@secondoftwo}%
\providecommand \href [0]{\begingroup \@sanitize@url \@href}%
\providecommand \@href[1]{\@@startlink{#1}\@@href}%
\providecommand \@@href[1]{\endgroup#1\@@endlink}%
\providecommand \@sanitize@url [0]{\catcode `\\12\catcode `\$12\catcode
  `\&12\catcode `\#12\catcode `\^12\catcode `\_12\catcode `\%12\relax}%
\providecommand \@@startlink[1]{}%
\providecommand \@@endlink[0]{}%
\providecommand \url  [0]{\begingroup\@sanitize@url \@url }%
\providecommand \@url [1]{\endgroup\@href {#1}{\urlprefix }}%
\providecommand \urlprefix  [0]{URL }%
\providecommand \Eprint [0]{\href }%
\providecommand \doibase [0]{http://dx.doi.org/}%
\providecommand \selectlanguage [0]{\@gobble}%
\providecommand \bibinfo  [0]{\@secondoftwo}%
\providecommand \bibfield  [0]{\@secondoftwo}%
\providecommand \translation [1]{[#1]}%
\providecommand \BibitemOpen [0]{}%
\providecommand \bibitemStop [0]{}%
\providecommand \bibitemNoStop [0]{.\EOS\space}%
\providecommand \EOS [0]{\spacefactor3000\relax}%
\providecommand \BibitemShut  [1]{\csname bibitem#1\endcsname}%
\let\auto@bib@innerbib\@empty
\bibitem [{\citenamefont {Arcadi}\ \emph {et~al.}(2018)\citenamefont {Arcadi},
  \citenamefont {Dutra}, \citenamefont {Ghosh}, \citenamefont {Lindner},
  \citenamefont {Mambrini}, \citenamefont {Pierre}, \citenamefont {Profumo},\
  and\ \citenamefont {Queiroz}}]{Arcadi:2017kky}%
  \BibitemOpen
  \bibfield  {author} {\bibinfo {author} {\bibfnamefont {G.}~\bibnamefont
  {Arcadi}}, \bibinfo {author} {\bibfnamefont {M.}~\bibnamefont {Dutra}},
  \bibinfo {author} {\bibfnamefont {P.}~\bibnamefont {Ghosh}}, \bibinfo
  {author} {\bibfnamefont {M.}~\bibnamefont {Lindner}}, \bibinfo {author}
  {\bibfnamefont {Y.}~\bibnamefont {Mambrini}}, \bibinfo {author}
  {\bibfnamefont {M.}~\bibnamefont {Pierre}}, \bibinfo {author} {\bibfnamefont
  {S.}~\bibnamefont {Profumo}}, \ and\ \bibinfo {author} {\bibfnamefont
  {F.~S.}\ \bibnamefont {Queiroz}},\ }\href {\doibase
  10.1140/epjc/s10052-018-5662-y} {\bibfield  {journal} {\bibinfo  {journal}
  {Eur. Phys. J. C}\ }\textbf {\bibinfo {volume} {78}},\ \bibinfo {pages} {203}
  (\bibinfo {year} {2018})},\ \Eprint {http://arxiv.org/abs/1703.07364}
  {arXiv:1703.07364 [hep-ph]} \BibitemShut {NoStop}%
\bibitem [{\citenamefont {Abbott}\ \emph {et~al.}(2016)\citenamefont {Abbott}
  \emph {et~al.}}]{LIGOScientific:2016aoc}%
  \BibitemOpen
  \bibfield  {author} {\bibinfo {author} {\bibfnamefont {B.~P.}\ \bibnamefont
  {Abbott}} \emph {et~al.} (\bibinfo {collaboration} {LIGO Scientific,
  Virgo}),\ }\href {\doibase 10.1103/PhysRevLett.116.061102} {\bibfield
  {journal} {\bibinfo  {journal} {Phys. Rev. Lett.}\ }\textbf {\bibinfo
  {volume} {116}},\ \bibinfo {pages} {061102} (\bibinfo {year} {2016})},\
  \Eprint {http://arxiv.org/abs/1602.03837} {arXiv:1602.03837 [gr-qc]}
  \BibitemShut {NoStop}%
\bibitem [{\citenamefont {Abbott}\ \emph {et~al.}(2019)\citenamefont {Abbott}
  \emph {et~al.}}]{LIGOScientific:2018mvr}%
  \BibitemOpen
  \bibfield  {author} {\bibinfo {author} {\bibfnamefont {B.~P.}\ \bibnamefont
  {Abbott}} \emph {et~al.} (\bibinfo {collaboration} {LIGO Scientific,
  Virgo}),\ }\href {\doibase 10.1103/PhysRevX.9.031040} {\bibfield  {journal}
  {\bibinfo  {journal} {Phys. Rev. X}\ }\textbf {\bibinfo {volume} {9}},\
  \bibinfo {pages} {031040} (\bibinfo {year} {2019})},\ \Eprint
  {http://arxiv.org/abs/1811.12907} {arXiv:1811.12907 [astro-ph.HE]}
  \BibitemShut {NoStop}%
\bibitem [{\citenamefont {Abbott}\ \emph
  {et~al.}(2021{\natexlab{a}})\citenamefont {Abbott} \emph
  {et~al.}}]{LIGOScientific:2021usb}%
  \BibitemOpen
  \bibfield  {author} {\bibinfo {author} {\bibfnamefont {R.}~\bibnamefont
  {Abbott}} \emph {et~al.} (\bibinfo {collaboration} {LIGO Scientific,
  VIRGO}),\ }\href@noop {} {\  (\bibinfo {year} {2021}{\natexlab{a}})},\
  \Eprint {http://arxiv.org/abs/2108.01045} {arXiv:2108.01045 [gr-qc]}
  \BibitemShut {NoStop}%
\bibitem [{\citenamefont {Abbott}\ \emph
  {et~al.}(2021{\natexlab{b}})\citenamefont {Abbott} \emph
  {et~al.}}]{LIGOScientific:2021djp}%
  \BibitemOpen
  \bibfield  {author} {\bibinfo {author} {\bibfnamefont {R.}~\bibnamefont
  {Abbott}} \emph {et~al.} (\bibinfo {collaboration} {LIGO Scientific, VIRGO,
  KAGRA}),\ }\href@noop {} {\  (\bibinfo {year} {2021}{\natexlab{b}})},\
  \Eprint {http://arxiv.org/abs/2111.03606} {arXiv:2111.03606 [gr-qc]}
  \BibitemShut {NoStop}%
\bibitem [{\citenamefont {Agazie}\ \emph {et~al.}(2023)\citenamefont {Agazie}
  \emph {et~al.}}]{NANOGrav:2023gor}%
  \BibitemOpen
  \bibfield  {author} {\bibinfo {author} {\bibfnamefont {G.}~\bibnamefont
  {Agazie}} \emph {et~al.} (\bibinfo {collaboration} {NANOGrav}),\ }\href
  {\doibase 10.3847/2041-8213/acdac6} {\bibfield  {journal} {\bibinfo
  {journal} {Astrophys. J. Lett.}\ }\textbf {\bibinfo {volume} {951}},\
  \bibinfo {pages} {L8} (\bibinfo {year} {2023})},\ \Eprint
  {http://arxiv.org/abs/2306.16213} {arXiv:2306.16213 [astro-ph.HE]}
  \BibitemShut {NoStop}%
\bibitem [{\citenamefont {Amaro-Seoane}\ \emph {et~al.}(2017)\citenamefont
  {Amaro-Seoane} \emph {et~al.}}]{LISA:2017pwj}%
  \BibitemOpen
  \bibfield  {author} {\bibinfo {author} {\bibfnamefont {P.}~\bibnamefont
  {Amaro-Seoane}} \emph {et~al.} (\bibinfo {collaboration} {LISA}),\
  }\href@noop {} {\  (\bibinfo {year} {2017})},\ \Eprint
  {http://arxiv.org/abs/1702.00786} {arXiv:1702.00786 [astro-ph.IM]}
  \BibitemShut {NoStop}%
\bibitem [{\citenamefont {Crowder}\ and\ \citenamefont
  {Cornish}(2005)}]{Crowder:2005nr}%
  \BibitemOpen
  \bibfield  {author} {\bibinfo {author} {\bibfnamefont {J.}~\bibnamefont
  {Crowder}}\ and\ \bibinfo {author} {\bibfnamefont {N.~J.}\ \bibnamefont
  {Cornish}},\ }\href {\doibase 10.1103/PhysRevD.72.083005} {\bibfield
  {journal} {\bibinfo  {journal} {Phys. Rev. D}\ }\textbf {\bibinfo {volume}
  {72}},\ \bibinfo {pages} {083005} (\bibinfo {year} {2005})},\ \Eprint
  {http://arxiv.org/abs/gr-qc/0506015} {arXiv:gr-qc/0506015} \BibitemShut
  {NoStop}%
\bibitem [{\citenamefont {Kawamura}\ \emph {et~al.}(2006)\citenamefont
  {Kawamura} \emph {et~al.}}]{Kawamura:2006up}%
  \BibitemOpen
  \bibfield  {author} {\bibinfo {author} {\bibfnamefont {S.}~\bibnamefont
  {Kawamura}} \emph {et~al.},\ }\href {\doibase 10.1088/0264-9381/23/8/S17}
  {\bibfield  {journal} {\bibinfo  {journal} {Class. Quant. Grav.}\ }\textbf
  {\bibinfo {volume} {23}},\ \bibinfo {pages} {S125} (\bibinfo {year}
  {2006})}\BibitemShut {NoStop}%
\bibitem [{\citenamefont {Aprile}\ \emph {et~al.}(2023)\citenamefont {Aprile}
  \emph {et~al.}}]{XENON:2023sxq}%
  \BibitemOpen
  \bibfield  {author} {\bibinfo {author} {\bibfnamefont {E.}~\bibnamefont
  {Aprile}} \emph {et~al.} (\bibinfo {collaboration} {XENON}),\ }\href@noop {}
  {\  (\bibinfo {year} {2023})},\ \Eprint {http://arxiv.org/abs/2303.14729}
  {arXiv:2303.14729 [hep-ex]} \BibitemShut {NoStop}%
\bibitem [{\citenamefont {Aalbers}\ \emph {et~al.}(2022)\citenamefont {Aalbers}
  \emph {et~al.}}]{LZ:2022ufs}%
  \BibitemOpen
  \bibfield  {author} {\bibinfo {author} {\bibfnamefont {J.}~\bibnamefont
  {Aalbers}} \emph {et~al.} (\bibinfo {collaboration} {LZ}),\ }\href@noop {} {\
   (\bibinfo {year} {2022})},\ \Eprint {http://arxiv.org/abs/2207.03764}
  {arXiv:2207.03764 [hep-ex]} \BibitemShut {NoStop}%
\bibitem [{\citenamefont {Aad}\ \emph {et~al.}(2019)\citenamefont {Aad} \emph
  {et~al.}}]{ATLAS:2019erb}%
  \BibitemOpen
  \bibfield  {author} {\bibinfo {author} {\bibfnamefont {G.}~\bibnamefont
  {Aad}} \emph {et~al.} (\bibinfo {collaboration} {ATLAS}),\ }\href {\doibase
  10.1016/j.physletb.2019.07.016} {\bibfield  {journal} {\bibinfo  {journal}
  {Phys. Lett. B}\ }\textbf {\bibinfo {volume} {796}},\ \bibinfo {pages} {68}
  (\bibinfo {year} {2019})},\ \Eprint {http://arxiv.org/abs/1903.06248}
  {arXiv:1903.06248 [hep-ex]} \BibitemShut {NoStop}%
\bibitem [{\citenamefont {Aad}\ \emph {et~al.}(2020)\citenamefont {Aad} \emph
  {et~al.}}]{ATLAS:2019fgd}%
  \BibitemOpen
  \bibfield  {author} {\bibinfo {author} {\bibfnamefont {G.}~\bibnamefont
  {Aad}} \emph {et~al.} (\bibinfo {collaboration} {ATLAS}),\ }\href {\doibase
  10.1007/JHEP03(2020)145} {\bibfield  {journal} {\bibinfo  {journal} {JHEP}\
  }\textbf {\bibinfo {volume} {03}},\ \bibinfo {pages} {145} (\bibinfo {year}
  {2020})},\ \Eprint {http://arxiv.org/abs/1910.08447} {arXiv:1910.08447
  [hep-ex]} \BibitemShut {NoStop}%
\bibitem [{\citenamefont {Aaboud}\ \emph {et~al.}(2019)\citenamefont {Aaboud}
  \emph {et~al.}}]{ATLAS:2019npw}%
  \BibitemOpen
  \bibfield  {author} {\bibinfo {author} {\bibfnamefont {M.}~\bibnamefont
  {Aaboud}} \emph {et~al.} (\bibinfo {collaboration} {ATLAS}),\ }\href
  {\doibase 10.1103/PhysRevD.99.092004} {\bibfield  {journal} {\bibinfo
  {journal} {Phys. Rev. D}\ }\textbf {\bibinfo {volume} {99}},\ \bibinfo
  {pages} {092004} (\bibinfo {year} {2019})},\ \Eprint
  {http://arxiv.org/abs/1902.10077} {arXiv:1902.10077 [hep-ex]} \BibitemShut
  {NoStop}%
\bibitem [{\citenamefont {Hasegawa}\ \emph {et~al.}(2019)\citenamefont
  {Hasegawa}, \citenamefont {Okada},\ and\ \citenamefont
  {Seto}}]{Hasegawa:2019amx}%
  \BibitemOpen
  \bibfield  {author} {\bibinfo {author} {\bibfnamefont {T.}~\bibnamefont
  {Hasegawa}}, \bibinfo {author} {\bibfnamefont {N.}~\bibnamefont {Okada}}, \
  and\ \bibinfo {author} {\bibfnamefont {O.}~\bibnamefont {Seto}},\ }\href
  {\doibase 10.1103/PhysRevD.99.095039} {\bibfield  {journal} {\bibinfo
  {journal} {Phys. Rev. D}\ }\textbf {\bibinfo {volume} {99}},\ \bibinfo
  {pages} {095039} (\bibinfo {year} {2019})},\ \Eprint
  {http://arxiv.org/abs/1904.03020} {arXiv:1904.03020 [hep-ph]} \BibitemShut
  {NoStop}%
\bibitem [{\citenamefont {Bosch}\ \emph {et~al.}(2023)\citenamefont {Bosch},
  \citenamefont {Delgado}, \citenamefont {Fornal},\ and\ \citenamefont
  {Leon}}]{Bosch:2023spa}%
  \BibitemOpen
  \bibfield  {author} {\bibinfo {author} {\bibfnamefont {J.}~\bibnamefont
  {Bosch}}, \bibinfo {author} {\bibfnamefont {Z.}~\bibnamefont {Delgado}},
  \bibinfo {author} {\bibfnamefont {B.}~\bibnamefont {Fornal}}, \ and\ \bibinfo
  {author} {\bibfnamefont {A.}~\bibnamefont {Leon}},\ }\href@noop {} {\
  (\bibinfo {year} {2023})},\ \Eprint {http://arxiv.org/abs/2306.00332}
  {arXiv:2306.00332 [hep-ph]} \BibitemShut {NoStop}%
\bibitem [{\citenamefont {Okada}\ and\ \citenamefont
  {Okada}(2016)}]{Okada:2016gsh}%
  \BibitemOpen
  \bibfield  {author} {\bibinfo {author} {\bibfnamefont {N.}~\bibnamefont
  {Okada}}\ and\ \bibinfo {author} {\bibfnamefont {S.}~\bibnamefont {Okada}},\
  }\href {\doibase 10.1103/PhysRevD.93.075003} {\bibfield  {journal} {\bibinfo
  {journal} {Phys. Rev. D}\ }\textbf {\bibinfo {volume} {93}},\ \bibinfo
  {pages} {075003} (\bibinfo {year} {2016})},\ \Eprint
  {http://arxiv.org/abs/1601.07526} {arXiv:1601.07526 [hep-ph]} \BibitemShut
  {NoStop}%
\bibitem [{\citenamefont {Rodejohann}\ and\ \citenamefont
  {Yaguna}(2015)}]{Rodejohann:2015lca}%
  \BibitemOpen
  \bibfield  {author} {\bibinfo {author} {\bibfnamefont {W.}~\bibnamefont
  {Rodejohann}}\ and\ \bibinfo {author} {\bibfnamefont {C.~E.}\ \bibnamefont
  {Yaguna}},\ }\href {\doibase 10.1088/1475-7516/2015/12/032} {\bibfield
  {journal} {\bibinfo  {journal} {JCAP}\ }\textbf {\bibinfo {volume} {12}},\
  \bibinfo {pages} {032} (\bibinfo {year} {2015})},\ \Eprint
  {http://arxiv.org/abs/1509.04036} {arXiv:1509.04036 [hep-ph]} \BibitemShut
  {NoStop}%
\bibitem [{\citenamefont {Madge}\ and\ \citenamefont
  {Schwaller}(2019)}]{Madge:2018gfl}%
  \BibitemOpen
  \bibfield  {author} {\bibinfo {author} {\bibfnamefont {E.}~\bibnamefont
  {Madge}}\ and\ \bibinfo {author} {\bibfnamefont {P.}~\bibnamefont
  {Schwaller}},\ }\href {\doibase 10.1007/JHEP02(2019)048} {\bibfield
  {journal} {\bibinfo  {journal} {JHEP}\ }\textbf {\bibinfo {volume} {02}},\
  \bibinfo {pages} {048} (\bibinfo {year} {2019})},\ \Eprint
  {http://arxiv.org/abs/1809.09110} {arXiv:1809.09110 [hep-ph]} \BibitemShut
  {NoStop}%
\bibitem [{\citenamefont {Breitbach}\ \emph {et~al.}(2019)\citenamefont
  {Breitbach}, \citenamefont {Kopp}, \citenamefont {Madge}, \citenamefont
  {Opferkuch},\ and\ \citenamefont {Schwaller}}]{Breitbach:2018ddu}%
  \BibitemOpen
  \bibfield  {author} {\bibinfo {author} {\bibfnamefont {M.}~\bibnamefont
  {Breitbach}}, \bibinfo {author} {\bibfnamefont {J.}~\bibnamefont {Kopp}},
  \bibinfo {author} {\bibfnamefont {E.}~\bibnamefont {Madge}}, \bibinfo
  {author} {\bibfnamefont {T.}~\bibnamefont {Opferkuch}}, \ and\ \bibinfo
  {author} {\bibfnamefont {P.}~\bibnamefont {Schwaller}},\ }\href {\doibase
  10.1088/1475-7516/2019/07/007} {\bibfield  {journal} {\bibinfo  {journal}
  {JCAP}\ }\textbf {\bibinfo {volume} {07}},\ \bibinfo {pages} {007} (\bibinfo
  {year} {2019})},\ \Eprint {http://arxiv.org/abs/1811.11175} {arXiv:1811.11175
  [hep-ph]} \BibitemShut {NoStop}%
\bibitem [{\citenamefont {Abe}\ and\ \citenamefont
  {Hashino}(2023)}]{Abe:2023zja}%
  \BibitemOpen
  \bibfield  {author} {\bibinfo {author} {\bibfnamefont {T.}~\bibnamefont
  {Abe}}\ and\ \bibinfo {author} {\bibfnamefont {K.}~\bibnamefont {Hashino}},\
  }\href@noop {} {\  (\bibinfo {year} {2023})},\ \Eprint
  {http://arxiv.org/abs/2302.13510} {arXiv:2302.13510 [hep-ph]} \BibitemShut
  {NoStop}%
\bibitem [{\citenamefont {Costa}\ \emph
  {et~al.}(2022{\natexlab{a}})\citenamefont {Costa}, \citenamefont {Khan},\
  and\ \citenamefont {Kim}}]{Costa:2022lpy}%
  \BibitemOpen
  \bibfield  {author} {\bibinfo {author} {\bibfnamefont {F.}~\bibnamefont
  {Costa}}, \bibinfo {author} {\bibfnamefont {S.}~\bibnamefont {Khan}}, \ and\
  \bibinfo {author} {\bibfnamefont {J.}~\bibnamefont {Kim}},\ }\href {\doibase
  10.1007/JHEP12(2022)165} {\bibfield  {journal} {\bibinfo  {journal} {JHEP}\
  }\textbf {\bibinfo {volume} {12}},\ \bibinfo {pages} {165} (\bibinfo {year}
  {2022}{\natexlab{a}})},\ \Eprint {http://arxiv.org/abs/2209.13653}
  {arXiv:2209.13653 [hep-ph]} \BibitemShut {NoStop}%
\bibitem [{\citenamefont {Costa}\ \emph
  {et~al.}(2022{\natexlab{b}})\citenamefont {Costa}, \citenamefont {Khan},\
  and\ \citenamefont {Kim}}]{Costa:2022oaa}%
  \BibitemOpen
  \bibfield  {author} {\bibinfo {author} {\bibfnamefont {F.}~\bibnamefont
  {Costa}}, \bibinfo {author} {\bibfnamefont {S.}~\bibnamefont {Khan}}, \ and\
  \bibinfo {author} {\bibfnamefont {J.}~\bibnamefont {Kim}},\ }\href {\doibase
  10.1007/JHEP06(2022)026} {\bibfield  {journal} {\bibinfo  {journal} {JHEP}\
  }\textbf {\bibinfo {volume} {06}},\ \bibinfo {pages} {026} (\bibinfo {year}
  {2022}{\natexlab{b}})},\ \Eprint {http://arxiv.org/abs/2202.13126}
  {arXiv:2202.13126 [hep-ph]} \BibitemShut {NoStop}%
\bibitem [{\citenamefont {Beniwal}\ \emph {et~al.}(2017)\citenamefont
  {Beniwal}, \citenamefont {Lewicki}, \citenamefont {Wells}, \citenamefont
  {White},\ and\ \citenamefont {Williams}}]{Beniwal:2017eik}%
  \BibitemOpen
  \bibfield  {author} {\bibinfo {author} {\bibfnamefont {A.}~\bibnamefont
  {Beniwal}}, \bibinfo {author} {\bibfnamefont {M.}~\bibnamefont {Lewicki}},
  \bibinfo {author} {\bibfnamefont {J.~D.}\ \bibnamefont {Wells}}, \bibinfo
  {author} {\bibfnamefont {M.}~\bibnamefont {White}}, \ and\ \bibinfo {author}
  {\bibfnamefont {A.~G.}\ \bibnamefont {Williams}},\ }\href {\doibase
  10.1007/JHEP08(2017)108} {\bibfield  {journal} {\bibinfo  {journal} {JHEP}\
  }\textbf {\bibinfo {volume} {08}},\ \bibinfo {pages} {108} (\bibinfo {year}
  {2017})},\ \Eprint {http://arxiv.org/abs/1702.06124} {arXiv:1702.06124
  [hep-ph]} \BibitemShut {NoStop}%
\bibitem [{\citenamefont {Arcadi}\ \emph {et~al.}(2022)\citenamefont {Arcadi},
  \citenamefont {Benincasa}, \citenamefont {Djouadi},\ and\ \citenamefont
  {Kannike}}]{Arcadi:2022lpp}%
  \BibitemOpen
  \bibfield  {author} {\bibinfo {author} {\bibfnamefont {G.}~\bibnamefont
  {Arcadi}}, \bibinfo {author} {\bibfnamefont {N.}~\bibnamefont {Benincasa}},
  \bibinfo {author} {\bibfnamefont {A.}~\bibnamefont {Djouadi}}, \ and\
  \bibinfo {author} {\bibfnamefont {K.}~\bibnamefont {Kannike}},\ }\href@noop
  {} {\  (\bibinfo {year} {2022})},\ \Eprint {http://arxiv.org/abs/2212.14788}
  {arXiv:2212.14788 [hep-ph]} \BibitemShut {NoStop}%
\bibitem [{\citenamefont {Klasen}\ \emph {et~al.}(2017)\citenamefont {Klasen},
  \citenamefont {Lyonnet},\ and\ \citenamefont {Queiroz}}]{Klasen:2016qux}%
  \BibitemOpen
  \bibfield  {author} {\bibinfo {author} {\bibfnamefont {M.}~\bibnamefont
  {Klasen}}, \bibinfo {author} {\bibfnamefont {F.}~\bibnamefont {Lyonnet}}, \
  and\ \bibinfo {author} {\bibfnamefont {F.~S.}\ \bibnamefont {Queiroz}},\
  }\href {\doibase 10.1140/epjc/s10052-017-4904-8} {\bibfield  {journal}
  {\bibinfo  {journal} {Eur. Phys. J. C}\ }\textbf {\bibinfo {volume} {77}},\
  \bibinfo {pages} {348} (\bibinfo {year} {2017})},\ \Eprint
  {http://arxiv.org/abs/1607.06468} {arXiv:1607.06468 [hep-ph]} \BibitemShut
  {NoStop}%
\bibitem [{\citenamefont {Camargo}\ \emph {et~al.}(2019)\citenamefont
  {Camargo}, \citenamefont {Delle~Rose}, \citenamefont {Moretti},\ and\
  \citenamefont {Queiroz}}]{Camargo:2018klg}%
  \BibitemOpen
  \bibfield  {author} {\bibinfo {author} {\bibfnamefont {D.~A.}\ \bibnamefont
  {Camargo}}, \bibinfo {author} {\bibfnamefont {L.}~\bibnamefont {Delle~Rose}},
  \bibinfo {author} {\bibfnamefont {S.}~\bibnamefont {Moretti}}, \ and\
  \bibinfo {author} {\bibfnamefont {F.~S.}\ \bibnamefont {Queiroz}},\ }\href
  {\doibase 10.1016/j.physletb.2019.04.048} {\bibfield  {journal} {\bibinfo
  {journal} {Phys. Lett. B}\ }\textbf {\bibinfo {volume} {793}},\ \bibinfo
  {pages} {150} (\bibinfo {year} {2019})},\ \Eprint
  {http://arxiv.org/abs/1805.08231} {arXiv:1805.08231 [hep-ph]} \BibitemShut
  {NoStop}%
\bibitem [{\citenamefont {Minkowski}(1977)}]{Minkowski:1977sc}%
  \BibitemOpen
  \bibfield  {author} {\bibinfo {author} {\bibfnamefont {P.}~\bibnamefont
  {Minkowski}},\ }\href {\doibase 10.1016/0370-2693(77)90435-X} {\bibfield
  {journal} {\bibinfo  {journal} {Phys. Lett. B}\ }\textbf {\bibinfo {volume}
  {67}},\ \bibinfo {pages} {421} (\bibinfo {year} {1977})}\BibitemShut
  {NoStop}%
\bibitem [{\citenamefont {Mohapatra}\ and\ \citenamefont
  {Senjanovic}(1980)}]{Mohapatra:1979ia}%
  \BibitemOpen
  \bibfield  {author} {\bibinfo {author} {\bibfnamefont {R.~N.}\ \bibnamefont
  {Mohapatra}}\ and\ \bibinfo {author} {\bibfnamefont {G.}~\bibnamefont
  {Senjanovic}},\ }\href {\doibase 10.1103/PhysRevLett.44.912} {\bibfield
  {journal} {\bibinfo  {journal} {Phys. Rev. Lett.}\ }\textbf {\bibinfo
  {volume} {44}},\ \bibinfo {pages} {912} (\bibinfo {year} {1980})}\BibitemShut
  {NoStop}%
\bibitem [{\citenamefont {Brdar}\ \emph {et~al.}(2019)\citenamefont {Brdar},
  \citenamefont {Helmboldt}, \citenamefont {Iwamoto},\ and\ \citenamefont
  {Schmitz}}]{Brdar:2019iem}%
  \BibitemOpen
  \bibfield  {author} {\bibinfo {author} {\bibfnamefont {V.}~\bibnamefont
  {Brdar}}, \bibinfo {author} {\bibfnamefont {A.~J.}\ \bibnamefont
  {Helmboldt}}, \bibinfo {author} {\bibfnamefont {S.}~\bibnamefont {Iwamoto}},
  \ and\ \bibinfo {author} {\bibfnamefont {K.}~\bibnamefont {Schmitz}},\ }\href
  {\doibase 10.1103/PhysRevD.100.075029} {\bibfield  {journal} {\bibinfo
  {journal} {Phys. Rev. D}\ }\textbf {\bibinfo {volume} {100}},\ \bibinfo
  {pages} {075029} (\bibinfo {year} {2019})},\ \Eprint
  {http://arxiv.org/abs/1905.12634} {arXiv:1905.12634 [hep-ph]} \BibitemShut
  {NoStop}%
\bibitem [{\citenamefont {Caprini}\ and\ \citenamefont
  {Figueroa}(2018)}]{Caprini:2018mtu}%
  \BibitemOpen
  \bibfield  {author} {\bibinfo {author} {\bibfnamefont {C.}~\bibnamefont
  {Caprini}}\ and\ \bibinfo {author} {\bibfnamefont {D.~G.}\ \bibnamefont
  {Figueroa}},\ }\href {\doibase 10.1088/1361-6382/aac608} {\bibfield
  {journal} {\bibinfo  {journal} {Class. Quant. Grav.}\ }\textbf {\bibinfo
  {volume} {35}},\ \bibinfo {pages} {163001} (\bibinfo {year} {2018})},\
  \Eprint {http://arxiv.org/abs/1801.04268} {arXiv:1801.04268 [astro-ph.CO]}
  \BibitemShut {NoStop}%
\bibitem [{\citenamefont {Caprini}\ \emph {et~al.}(2020)\citenamefont {Caprini}
  \emph {et~al.}}]{Caprini:2019egz}%
  \BibitemOpen
  \bibfield  {author} {\bibinfo {author} {\bibfnamefont {C.}~\bibnamefont
  {Caprini}} \emph {et~al.},\ }\href {\doibase 10.1088/1475-7516/2020/03/024}
  {\bibfield  {journal} {\bibinfo  {journal} {JCAP}\ }\textbf {\bibinfo
  {volume} {03}},\ \bibinfo {pages} {024} (\bibinfo {year} {2020})},\ \Eprint
  {http://arxiv.org/abs/1910.13125} {arXiv:1910.13125 [astro-ph.CO]}
  \BibitemShut {NoStop}%
\bibitem [{\citenamefont {Dolan}\ and\ \citenamefont
  {Jackiw}(1974)}]{Dolan:1973qd}%
  \BibitemOpen
  \bibfield  {author} {\bibinfo {author} {\bibfnamefont {L.}~\bibnamefont
  {Dolan}}\ and\ \bibinfo {author} {\bibfnamefont {R.}~\bibnamefont {Jackiw}},\
  }\href {\doibase 10.1103/PhysRevD.9.3320} {\bibfield  {journal} {\bibinfo
  {journal} {Phys. Rev. D}\ }\textbf {\bibinfo {volume} {9}},\ \bibinfo {pages}
  {3320} (\bibinfo {year} {1974})}\BibitemShut {NoStop}%
\bibitem [{\citenamefont {Hindmarsh}\ \emph {et~al.}(2021)\citenamefont
  {Hindmarsh}, \citenamefont {L\"uben}, \citenamefont {Lumma},\ and\
  \citenamefont {Pauly}}]{Hindmarsh:2020hop}%
  \BibitemOpen
  \bibfield  {author} {\bibinfo {author} {\bibfnamefont {M.~B.}\ \bibnamefont
  {Hindmarsh}}, \bibinfo {author} {\bibfnamefont {M.}~\bibnamefont {L\"uben}},
  \bibinfo {author} {\bibfnamefont {J.}~\bibnamefont {Lumma}}, \ and\ \bibinfo
  {author} {\bibfnamefont {M.}~\bibnamefont {Pauly}},\ }\href {\doibase
  10.21468/SciPostPhysLectNotes.24} {\bibfield  {journal} {\bibinfo  {journal}
  {SciPost Phys. Lect. Notes}\ }\textbf {\bibinfo {volume} {24}},\ \bibinfo
  {pages} {1} (\bibinfo {year} {2021})},\ \Eprint
  {http://arxiv.org/abs/2008.09136} {arXiv:2008.09136 [astro-ph.CO]}
  \BibitemShut {NoStop}%
\bibitem [{\citenamefont {Mukhanov}(2005)}]{Mukhanov:2005sc}%
  \BibitemOpen
  \bibfield  {author} {\bibinfo {author} {\bibfnamefont {V.}~\bibnamefont
  {Mukhanov}},\ }\href {\doibase 10.1017/CBO9780511790553} {\emph {\bibinfo
  {title} {{Physical Foundations of Cosmology}}}}\ (\bibinfo  {publisher}
  {Cambridge University Press},\ \bibinfo {address} {Oxford},\ \bibinfo {year}
  {2005})\BibitemShut {NoStop}%
\bibitem [{\citenamefont {Espinosa}\ \emph {et~al.}(2010)\citenamefont
  {Espinosa}, \citenamefont {Konstandin}, \citenamefont {No},\ and\
  \citenamefont {Servant}}]{Espinosa:2010hh}%
  \BibitemOpen
  \bibfield  {author} {\bibinfo {author} {\bibfnamefont {J.~R.}\ \bibnamefont
  {Espinosa}}, \bibinfo {author} {\bibfnamefont {T.}~\bibnamefont
  {Konstandin}}, \bibinfo {author} {\bibfnamefont {J.~M.}\ \bibnamefont {No}},
  \ and\ \bibinfo {author} {\bibfnamefont {G.}~\bibnamefont {Servant}},\ }\href
  {\doibase 10.1088/1475-7516/2010/06/028} {\bibfield  {journal} {\bibinfo
  {journal} {JCAP}\ }\textbf {\bibinfo {volume} {06}},\ \bibinfo {pages} {028}
  (\bibinfo {year} {2010})},\ \Eprint {http://arxiv.org/abs/1004.4187}
  {arXiv:1004.4187 [hep-ph]} \BibitemShut {NoStop}%
\bibitem [{\citenamefont {Giese}\ \emph {et~al.}(2020)\citenamefont {Giese},
  \citenamefont {Konstandin},\ and\ \citenamefont {van~de
  Vis}}]{Giese:2020rtr}%
  \BibitemOpen
  \bibfield  {author} {\bibinfo {author} {\bibfnamefont {F.}~\bibnamefont
  {Giese}}, \bibinfo {author} {\bibfnamefont {T.}~\bibnamefont {Konstandin}}, \
  and\ \bibinfo {author} {\bibfnamefont {J.}~\bibnamefont {van~de Vis}},\
  }\href {\doibase 10.1088/1475-7516/2020/07/057} {\bibfield  {journal}
  {\bibinfo  {journal} {JCAP}\ }\textbf {\bibinfo {volume} {07}},\ \bibinfo
  {pages} {057} (\bibinfo {year} {2020})},\ \Eprint
  {http://arxiv.org/abs/2004.06995} {arXiv:2004.06995 [astro-ph.CO]}
  \BibitemShut {NoStop}%
\bibitem [{\citenamefont {Caprini}\ \emph {et~al.}(2016)\citenamefont {Caprini}
  \emph {et~al.}}]{Caprini:2015zlo}%
  \BibitemOpen
  \bibfield  {author} {\bibinfo {author} {\bibfnamefont {C.}~\bibnamefont
  {Caprini}} \emph {et~al.},\ }\href {\doibase 10.1088/1475-7516/2016/04/001}
  {\bibfield  {journal} {\bibinfo  {journal} {JCAP}\ }\textbf {\bibinfo
  {volume} {04}},\ \bibinfo {pages} {001} (\bibinfo {year} {2016})},\ \Eprint
  {http://arxiv.org/abs/1512.06239} {arXiv:1512.06239 [astro-ph.CO]}
  \BibitemShut {NoStop}%
\bibitem [{\citenamefont {Guo}\ \emph {et~al.}(2021)\citenamefont {Guo},
  \citenamefont {Sinha}, \citenamefont {Vagie},\ and\ \citenamefont
  {White}}]{Guo:2020grp}%
  \BibitemOpen
  \bibfield  {author} {\bibinfo {author} {\bibfnamefont {H.-K.}\ \bibnamefont
  {Guo}}, \bibinfo {author} {\bibfnamefont {K.}~\bibnamefont {Sinha}}, \bibinfo
  {author} {\bibfnamefont {D.}~\bibnamefont {Vagie}}, \ and\ \bibinfo {author}
  {\bibfnamefont {G.}~\bibnamefont {White}},\ }\href {\doibase
  10.1088/1475-7516/2021/01/001} {\bibfield  {journal} {\bibinfo  {journal}
  {JCAP}\ }\textbf {\bibinfo {volume} {01}},\ \bibinfo {pages} {001} (\bibinfo
  {year} {2021})},\ \Eprint {http://arxiv.org/abs/2007.08537} {arXiv:2007.08537
  [hep-ph]} \BibitemShut {NoStop}%
\bibitem [{\citenamefont {Cline}\ and\ \citenamefont
  {Kainulainen}(2020)}]{Cline:2020jre}%
  \BibitemOpen
  \bibfield  {author} {\bibinfo {author} {\bibfnamefont {J.~M.}\ \bibnamefont
  {Cline}}\ and\ \bibinfo {author} {\bibfnamefont {K.}~\bibnamefont
  {Kainulainen}},\ }\href {\doibase 10.1103/PhysRevD.101.063525} {\bibfield
  {journal} {\bibinfo  {journal} {Phys. Rev. D}\ }\textbf {\bibinfo {volume}
  {101}},\ \bibinfo {pages} {063525} (\bibinfo {year} {2020})},\ \Eprint
  {http://arxiv.org/abs/2001.00568} {arXiv:2001.00568 [hep-ph]} \BibitemShut
  {NoStop}%
\bibitem [{\citenamefont {Dorsch}\ \emph {et~al.}(2022)\citenamefont {Dorsch},
  \citenamefont {Huber},\ and\ \citenamefont {Konstandin}}]{Dorsch:2021nje}%
  \BibitemOpen
  \bibfield  {author} {\bibinfo {author} {\bibfnamefont {G.~C.}\ \bibnamefont
  {Dorsch}}, \bibinfo {author} {\bibfnamefont {S.~J.}\ \bibnamefont {Huber}}, \
  and\ \bibinfo {author} {\bibfnamefont {T.}~\bibnamefont {Konstandin}},\
  }\href {\doibase 10.1088/1475-7516/2022/04/010} {\bibfield  {journal}
  {\bibinfo  {journal} {JCAP}\ }\textbf {\bibinfo {volume} {04}},\ \bibinfo
  {pages} {010} (\bibinfo {year} {2022})},\ \Eprint
  {http://arxiv.org/abs/2112.12548} {arXiv:2112.12548 [hep-ph]} \BibitemShut
  {NoStop}%
\bibitem [{\citenamefont {Laurent}\ and\ \citenamefont
  {Cline}(2022)}]{Laurent:2022jrs}%
  \BibitemOpen
  \bibfield  {author} {\bibinfo {author} {\bibfnamefont {B.}~\bibnamefont
  {Laurent}}\ and\ \bibinfo {author} {\bibfnamefont {J.~M.}\ \bibnamefont
  {Cline}},\ }\href {\doibase 10.1103/PhysRevD.106.023501} {\bibfield
  {journal} {\bibinfo  {journal} {Phys. Rev. D}\ }\textbf {\bibinfo {volume}
  {106}},\ \bibinfo {pages} {023501} (\bibinfo {year} {2022})},\ \Eprint
  {http://arxiv.org/abs/2204.13120} {arXiv:2204.13120 [hep-ph]} \BibitemShut
  {NoStop}%
\bibitem [{\citenamefont {Hindmarsh}\ \emph {et~al.}(2015)\citenamefont
  {Hindmarsh}, \citenamefont {Huber}, \citenamefont {Rummukainen},\ and\
  \citenamefont {Weir}}]{Hindmarsh:2015qta}%
  \BibitemOpen
  \bibfield  {author} {\bibinfo {author} {\bibfnamefont {M.}~\bibnamefont
  {Hindmarsh}}, \bibinfo {author} {\bibfnamefont {S.~J.}\ \bibnamefont
  {Huber}}, \bibinfo {author} {\bibfnamefont {K.}~\bibnamefont {Rummukainen}},
  \ and\ \bibinfo {author} {\bibfnamefont {D.~J.}\ \bibnamefont {Weir}},\
  }\href {\doibase 10.1103/PhysRevD.92.123009} {\bibfield  {journal} {\bibinfo
  {journal} {Phys. Rev. D}\ }\textbf {\bibinfo {volume} {92}},\ \bibinfo
  {pages} {123009} (\bibinfo {year} {2015})},\ \Eprint
  {http://arxiv.org/abs/1504.03291} {arXiv:1504.03291 [astro-ph.CO]}
  \BibitemShut {NoStop}%
\bibitem [{\citenamefont {Hindmarsh}\ \emph {et~al.}(2017)\citenamefont
  {Hindmarsh}, \citenamefont {Huber}, \citenamefont {Rummukainen},\ and\
  \citenamefont {Weir}}]{Hindmarsh:2017gnf}%
  \BibitemOpen
  \bibfield  {author} {\bibinfo {author} {\bibfnamefont {M.}~\bibnamefont
  {Hindmarsh}}, \bibinfo {author} {\bibfnamefont {S.~J.}\ \bibnamefont
  {Huber}}, \bibinfo {author} {\bibfnamefont {K.}~\bibnamefont {Rummukainen}},
  \ and\ \bibinfo {author} {\bibfnamefont {D.~J.}\ \bibnamefont {Weir}},\
  }\href {\doibase 10.1103/PhysRevD.96.103520} {\bibfield  {journal} {\bibinfo
  {journal} {Phys. Rev. D}\ }\textbf {\bibinfo {volume} {96}},\ \bibinfo
  {pages} {103520} (\bibinfo {year} {2017})},\ \bibinfo {note} {[Erratum:
  Phys.Rev.D 101, 089902 (2020)]},\ \Eprint {http://arxiv.org/abs/1704.05871}
  {arXiv:1704.05871 [astro-ph.CO]} \BibitemShut {NoStop}%
\bibitem [{\citenamefont {Aghanim}\ \emph {et~al.}(2020)\citenamefont {Aghanim}
  \emph {et~al.}}]{Planck:2018vyg}%
  \BibitemOpen
  \bibfield  {author} {\bibinfo {author} {\bibfnamefont {N.}~\bibnamefont
  {Aghanim}} \emph {et~al.} (\bibinfo {collaboration} {Planck}),\ }\href
  {\doibase 10.1051/0004-6361/201833910} {\bibfield  {journal} {\bibinfo
  {journal} {Astron. Astrophys.}\ }\textbf {\bibinfo {volume} {641}},\ \bibinfo
  {pages} {A6} (\bibinfo {year} {2020})},\ \bibinfo {note} {[Erratum:
  Astron.Astrophys. 652, C4 (2021)]},\ \Eprint
  {http://arxiv.org/abs/1807.06209} {arXiv:1807.06209 [astro-ph.CO]}
  \BibitemShut {NoStop}%
\bibitem [{\citenamefont {Belanger}\ \emph {et~al.}(2007)\citenamefont
  {Belanger}, \citenamefont {Boudjema}, \citenamefont {Pukhov},\ and\
  \citenamefont {Semenov}}]{Belanger:2006is}%
  \BibitemOpen
  \bibfield  {author} {\bibinfo {author} {\bibfnamefont {G.}~\bibnamefont
  {Belanger}}, \bibinfo {author} {\bibfnamefont {F.}~\bibnamefont {Boudjema}},
  \bibinfo {author} {\bibfnamefont {A.}~\bibnamefont {Pukhov}}, \ and\ \bibinfo
  {author} {\bibfnamefont {A.}~\bibnamefont {Semenov}},\ }\href {\doibase
  10.1016/j.cpc.2006.11.008} {\bibfield  {journal} {\bibinfo  {journal}
  {Comput. Phys. Commun.}\ }\textbf {\bibinfo {volume} {176}},\ \bibinfo
  {pages} {367} (\bibinfo {year} {2007})},\ \Eprint
  {http://arxiv.org/abs/hep-ph/0607059} {arXiv:hep-ph/0607059} \BibitemShut
  {NoStop}%
\bibitem [{\citenamefont {Belanger}\ \emph {et~al.}(2021)\citenamefont
  {Belanger}, \citenamefont {Mjallal},\ and\ \citenamefont
  {Pukhov}}]{Belanger:2020gnr}%
  \BibitemOpen
  \bibfield  {author} {\bibinfo {author} {\bibfnamefont {G.}~\bibnamefont
  {Belanger}}, \bibinfo {author} {\bibfnamefont {A.}~\bibnamefont {Mjallal}}, \
  and\ \bibinfo {author} {\bibfnamefont {A.}~\bibnamefont {Pukhov}},\ }\href
  {\doibase 10.1140/epjc/s10052-021-09012-z} {\bibfield  {journal} {\bibinfo
  {journal} {Eur. Phys. J. C}\ }\textbf {\bibinfo {volume} {81}},\ \bibinfo
  {pages} {239} (\bibinfo {year} {2021})},\ \Eprint
  {http://arxiv.org/abs/2003.08621} {arXiv:2003.08621 [hep-ph]} \BibitemShut
  {NoStop}%
\bibitem [{\citenamefont {Arcadi}\ \emph {et~al.}(2014)\citenamefont {Arcadi},
  \citenamefont {Mambrini}, \citenamefont {Tytgat},\ and\ \citenamefont
  {Zaldivar}}]{Arcadi:2013qia}%
  \BibitemOpen
  \bibfield  {author} {\bibinfo {author} {\bibfnamefont {G.}~\bibnamefont
  {Arcadi}}, \bibinfo {author} {\bibfnamefont {Y.}~\bibnamefont {Mambrini}},
  \bibinfo {author} {\bibfnamefont {M.~H.~G.}\ \bibnamefont {Tytgat}}, \ and\
  \bibinfo {author} {\bibfnamefont {B.}~\bibnamefont {Zaldivar}},\ }\href
  {\doibase 10.1007/JHEP03(2014)134} {\bibfield  {journal} {\bibinfo  {journal}
  {JHEP}\ }\textbf {\bibinfo {volume} {03}},\ \bibinfo {pages} {134} (\bibinfo
  {year} {2014})},\ \Eprint {http://arxiv.org/abs/1401.0221} {arXiv:1401.0221
  [hep-ph]} \BibitemShut {NoStop}%
\bibitem [{\citenamefont {Belyaev}\ \emph {et~al.}(2013)\citenamefont
  {Belyaev}, \citenamefont {Christensen},\ and\ \citenamefont
  {Pukhov}}]{Belyaev:2012qa}%
  \BibitemOpen
  \bibfield  {author} {\bibinfo {author} {\bibfnamefont {A.}~\bibnamefont
  {Belyaev}}, \bibinfo {author} {\bibfnamefont {N.~D.}\ \bibnamefont
  {Christensen}}, \ and\ \bibinfo {author} {\bibfnamefont {A.}~\bibnamefont
  {Pukhov}},\ }\href {\doibase 10.1016/j.cpc.2013.01.014} {\bibfield  {journal}
  {\bibinfo  {journal} {Comput. Phys. Commun.}\ }\textbf {\bibinfo {volume}
  {184}},\ \bibinfo {pages} {1729} (\bibinfo {year} {2013})},\ \Eprint
  {http://arxiv.org/abs/1207.6082} {arXiv:1207.6082 [hep-ph]} \BibitemShut
  {NoStop}%
\bibitem [{\citenamefont {Alwall}\ \emph {et~al.}(2014)\citenamefont {Alwall},
  \citenamefont {Frederix}, \citenamefont {Frixione}, \citenamefont {Hirschi},
  \citenamefont {Maltoni}, \citenamefont {Mattelaer}, \citenamefont {Shao},
  \citenamefont {Stelzer}, \citenamefont {Torrielli},\ and\ \citenamefont
  {Zaro}}]{Alwall:2014hca}%
  \BibitemOpen
  \bibfield  {author} {\bibinfo {author} {\bibfnamefont {J.}~\bibnamefont
  {Alwall}}, \bibinfo {author} {\bibfnamefont {R.}~\bibnamefont {Frederix}},
  \bibinfo {author} {\bibfnamefont {S.}~\bibnamefont {Frixione}}, \bibinfo
  {author} {\bibfnamefont {V.}~\bibnamefont {Hirschi}}, \bibinfo {author}
  {\bibfnamefont {F.}~\bibnamefont {Maltoni}}, \bibinfo {author} {\bibfnamefont
  {O.}~\bibnamefont {Mattelaer}}, \bibinfo {author} {\bibfnamefont {H.~S.}\
  \bibnamefont {Shao}}, \bibinfo {author} {\bibfnamefont {T.}~\bibnamefont
  {Stelzer}}, \bibinfo {author} {\bibfnamefont {P.}~\bibnamefont {Torrielli}},
  \ and\ \bibinfo {author} {\bibfnamefont {M.}~\bibnamefont {Zaro}},\ }\href
  {\doibase 10.1007/JHEP07(2014)079} {\bibfield  {journal} {\bibinfo  {journal}
  {JHEP}\ }\textbf {\bibinfo {volume} {07}},\ \bibinfo {pages} {079} (\bibinfo
  {year} {2014})},\ \Eprint {http://arxiv.org/abs/1405.0301} {arXiv:1405.0301
  [hep-ph]} \BibitemShut {NoStop}%
\bibitem [{\citenamefont {Frederix}\ \emph {et~al.}(2018)\citenamefont
  {Frederix}, \citenamefont {Frixione}, \citenamefont {Hirschi}, \citenamefont
  {Pagani}, \citenamefont {Shao},\ and\ \citenamefont
  {Zaro}}]{Frederix:2018nkq}%
  \BibitemOpen
  \bibfield  {author} {\bibinfo {author} {\bibfnamefont {R.}~\bibnamefont
  {Frederix}}, \bibinfo {author} {\bibfnamefont {S.}~\bibnamefont {Frixione}},
  \bibinfo {author} {\bibfnamefont {V.}~\bibnamefont {Hirschi}}, \bibinfo
  {author} {\bibfnamefont {D.}~\bibnamefont {Pagani}}, \bibinfo {author}
  {\bibfnamefont {H.~S.}\ \bibnamefont {Shao}}, \ and\ \bibinfo {author}
  {\bibfnamefont {M.}~\bibnamefont {Zaro}},\ }\href {\doibase
  10.1007/JHEP11(2021)085} {\bibfield  {journal} {\bibinfo  {journal} {JHEP}\
  }\textbf {\bibinfo {volume} {07}},\ \bibinfo {pages} {185} (\bibinfo {year}
  {2018})},\ \bibinfo {note} {[Erratum: JHEP 11, 085 (2021)]},\ \Eprint
  {http://arxiv.org/abs/1804.10017} {arXiv:1804.10017 [hep-ph]} \BibitemShut
  {NoStop}%
\bibitem [{\citenamefont {Carrazza}\ \emph {et~al.}(2013)\citenamefont
  {Carrazza}, \citenamefont {Forte},\ and\ \citenamefont
  {Rojo}}]{Carrazza:2013axa}%
  \BibitemOpen
  \bibfield  {author} {\bibinfo {author} {\bibfnamefont {S.}~\bibnamefont
  {Carrazza}}, \bibinfo {author} {\bibfnamefont {S.}~\bibnamefont {Forte}}, \
  and\ \bibinfo {author} {\bibfnamefont {J.}~\bibnamefont {Rojo}},\ }in\
  \href@noop {} {\emph {\bibinfo {booktitle} {{43rd International Symposium on
  Multiparticle Dynamics}}}}\ (\bibinfo {year} {2013})\ pp.\ \bibinfo {pages}
  {89--96},\ \Eprint {http://arxiv.org/abs/1311.5887} {arXiv:1311.5887
  [hep-ph]} \BibitemShut {NoStop}%
\bibitem [{\citenamefont {Thamm}\ \emph {et~al.}(2015)\citenamefont {Thamm},
  \citenamefont {Torre},\ and\ \citenamefont {Wulzer}}]{Thamm:2015zwa}%
  \BibitemOpen
  \bibfield  {author} {\bibinfo {author} {\bibfnamefont {A.}~\bibnamefont
  {Thamm}}, \bibinfo {author} {\bibfnamefont {R.}~\bibnamefont {Torre}}, \ and\
  \bibinfo {author} {\bibfnamefont {A.}~\bibnamefont {Wulzer}},\ }\href
  {\doibase 10.1007/JHEP07(2015)100} {\bibfield  {journal} {\bibinfo  {journal}
  {JHEP}\ }\textbf {\bibinfo {volume} {07}},\ \bibinfo {pages} {100} (\bibinfo
  {year} {2015})},\ \Eprint {http://arxiv.org/abs/1502.01701} {arXiv:1502.01701
  [hep-ph]} \BibitemShut {NoStop}%
\bibitem [{\citenamefont {Carena}\ \emph {et~al.}(2004)\citenamefont {Carena},
  \citenamefont {Daleo}, \citenamefont {Dobrescu},\ and\ \citenamefont
  {Tait}}]{Carena:2004xs}%
  \BibitemOpen
  \bibfield  {author} {\bibinfo {author} {\bibfnamefont {M.}~\bibnamefont
  {Carena}}, \bibinfo {author} {\bibfnamefont {A.}~\bibnamefont {Daleo}},
  \bibinfo {author} {\bibfnamefont {B.~A.}\ \bibnamefont {Dobrescu}}, \ and\
  \bibinfo {author} {\bibfnamefont {T.~M.~P.}\ \bibnamefont {Tait}},\ }\href
  {\doibase 10.1103/PhysRevD.70.093009} {\bibfield  {journal} {\bibinfo
  {journal} {Phys. Rev. D}\ }\textbf {\bibinfo {volume} {70}},\ \bibinfo
  {pages} {093009} (\bibinfo {year} {2004})},\ \Eprint
  {http://arxiv.org/abs/hep-ph/0408098} {arXiv:hep-ph/0408098} \BibitemShut
  {NoStop}%
\bibitem [{\citenamefont {Cacciapaglia}\ \emph {et~al.}(2006)\citenamefont
  {Cacciapaglia}, \citenamefont {Csaki}, \citenamefont {Marandella},\ and\
  \citenamefont {Strumia}}]{Cacciapaglia:2006pk}%
  \BibitemOpen
  \bibfield  {author} {\bibinfo {author} {\bibfnamefont {G.}~\bibnamefont
  {Cacciapaglia}}, \bibinfo {author} {\bibfnamefont {C.}~\bibnamefont {Csaki}},
  \bibinfo {author} {\bibfnamefont {G.}~\bibnamefont {Marandella}}, \ and\
  \bibinfo {author} {\bibfnamefont {A.}~\bibnamefont {Strumia}},\ }\href
  {\doibase 10.1103/PhysRevD.74.033011} {\bibfield  {journal} {\bibinfo
  {journal} {Phys. Rev. D}\ }\textbf {\bibinfo {volume} {74}},\ \bibinfo
  {pages} {033011} (\bibinfo {year} {2006})},\ \Eprint
  {http://arxiv.org/abs/hep-ph/0604111} {arXiv:hep-ph/0604111} \BibitemShut
  {NoStop}%
\bibitem [{\citenamefont {Ackermann}\ \emph {et~al.}(2015)\citenamefont
  {Ackermann} \emph {et~al.}}]{Fermi-LAT:2015att}%
  \BibitemOpen
  \bibfield  {author} {\bibinfo {author} {\bibfnamefont {M.}~\bibnamefont
  {Ackermann}} \emph {et~al.} (\bibinfo {collaboration} {Fermi-LAT}),\ }\href
  {\doibase 10.1103/PhysRevLett.115.231301} {\bibfield  {journal} {\bibinfo
  {journal} {Phys. Rev. Lett.}\ }\textbf {\bibinfo {volume} {115}},\ \bibinfo
  {pages} {231301} (\bibinfo {year} {2015})},\ \Eprint
  {http://arxiv.org/abs/1503.02641} {arXiv:1503.02641 [astro-ph.HE]}
  \BibitemShut {NoStop}%
\bibitem [{\citenamefont {Acharyya}\ \emph {et~al.}(2021)\citenamefont
  {Acharyya} \emph {et~al.}}]{CTA:2020qlo}%
  \BibitemOpen
  \bibfield  {author} {\bibinfo {author} {\bibfnamefont {A.}~\bibnamefont
  {Acharyya}} \emph {et~al.} (\bibinfo {collaboration} {CTA}),\ }\href
  {\doibase 10.1088/1475-7516/2021/01/057} {\bibfield  {journal} {\bibinfo
  {journal} {JCAP}\ }\textbf {\bibinfo {volume} {01}},\ \bibinfo {pages} {057}
  (\bibinfo {year} {2021})},\ \Eprint {http://arxiv.org/abs/2007.16129}
  {arXiv:2007.16129 [astro-ph.HE]} \BibitemShut {NoStop}%
\bibitem [{\citenamefont {Hooper}\ \emph {et~al.}(2013)\citenamefont {Hooper},
  \citenamefont {Kelso},\ and\ \citenamefont {Queiroz}}]{Hooper:2012sr}%
  \BibitemOpen
  \bibfield  {author} {\bibinfo {author} {\bibfnamefont {D.}~\bibnamefont
  {Hooper}}, \bibinfo {author} {\bibfnamefont {C.}~\bibnamefont {Kelso}}, \
  and\ \bibinfo {author} {\bibfnamefont {F.~S.}\ \bibnamefont {Queiroz}},\
  }\href {\doibase 10.1016/j.astropartphys.2013.04.007} {\bibfield  {journal}
  {\bibinfo  {journal} {Astropart. Phys.}\ }\textbf {\bibinfo {volume} {46}},\
  \bibinfo {pages} {55} (\bibinfo {year} {2013})},\ \Eprint
  {http://arxiv.org/abs/1209.3015} {arXiv:1209.3015 [astro-ph.HE]} \BibitemShut
  {NoStop}%
\bibitem [{\citenamefont {Acharya}\ \emph {et~al.}(2018)\citenamefont {Acharya}
  \emph {et~al.}}]{CTAConsortium:2017dvg}%
  \BibitemOpen
  \bibfield  {author} {\bibinfo {author} {\bibfnamefont {B.~S.}\ \bibnamefont
  {Acharya}} \emph {et~al.} (\bibinfo {collaboration} {CTA Consortium}),\
  }\href {\doibase 10.1142/10986} {\emph {\bibinfo {title} {{Science with the
  Cherenkov Telescope Array}}}}\ (\bibinfo  {publisher} {WSP},\ \bibinfo {year}
  {2018})\ \Eprint {http://arxiv.org/abs/1709.07997} {arXiv:1709.07997
  [astro-ph.IM]} \BibitemShut {NoStop}%
\bibitem [{\citenamefont {Alves}\ \emph {et~al.}(2014)\citenamefont {Alves},
  \citenamefont {Profumo},\ and\ \citenamefont {Queiroz}}]{Alves:2013tqa}%
  \BibitemOpen
  \bibfield  {author} {\bibinfo {author} {\bibfnamefont {A.}~\bibnamefont
  {Alves}}, \bibinfo {author} {\bibfnamefont {S.}~\bibnamefont {Profumo}}, \
  and\ \bibinfo {author} {\bibfnamefont {F.~S.}\ \bibnamefont {Queiroz}},\
  }\href {\doibase 10.1007/JHEP04(2014)063} {\bibfield  {journal} {\bibinfo
  {journal} {JHEP}\ }\textbf {\bibinfo {volume} {04}},\ \bibinfo {pages} {063}
  (\bibinfo {year} {2014})},\ \Eprint {http://arxiv.org/abs/1312.5281}
  {arXiv:1312.5281 [hep-ph]} \BibitemShut {NoStop}%
\bibitem [{\citenamefont {Alves}\ \emph
  {et~al.}(2015{\natexlab{a}})\citenamefont {Alves}, \citenamefont {Berlin},
  \citenamefont {Profumo},\ and\ \citenamefont {Queiroz}}]{Alves:2015pea}%
  \BibitemOpen
  \bibfield  {author} {\bibinfo {author} {\bibfnamefont {A.}~\bibnamefont
  {Alves}}, \bibinfo {author} {\bibfnamefont {A.}~\bibnamefont {Berlin}},
  \bibinfo {author} {\bibfnamefont {S.}~\bibnamefont {Profumo}}, \ and\
  \bibinfo {author} {\bibfnamefont {F.~S.}\ \bibnamefont {Queiroz}},\ }\href
  {\doibase 10.1103/PhysRevD.92.083004} {\bibfield  {journal} {\bibinfo
  {journal} {Phys. Rev. D}\ }\textbf {\bibinfo {volume} {92}},\ \bibinfo
  {pages} {083004} (\bibinfo {year} {2015}{\natexlab{a}})},\ \Eprint
  {http://arxiv.org/abs/1501.03490} {arXiv:1501.03490 [hep-ph]} \BibitemShut
  {NoStop}%
\bibitem [{\citenamefont {Alves}\ \emph
  {et~al.}(2015{\natexlab{b}})\citenamefont {Alves}, \citenamefont {Berlin},
  \citenamefont {Profumo},\ and\ \citenamefont {Queiroz}}]{Alves:2015mua}%
  \BibitemOpen
  \bibfield  {author} {\bibinfo {author} {\bibfnamefont {A.}~\bibnamefont
  {Alves}}, \bibinfo {author} {\bibfnamefont {A.}~\bibnamefont {Berlin}},
  \bibinfo {author} {\bibfnamefont {S.}~\bibnamefont {Profumo}}, \ and\
  \bibinfo {author} {\bibfnamefont {F.~S.}\ \bibnamefont {Queiroz}},\ }\href
  {\doibase 10.1007/JHEP10(2015)076} {\bibfield  {journal} {\bibinfo  {journal}
  {JHEP}\ }\textbf {\bibinfo {volume} {10}},\ \bibinfo {pages} {076} (\bibinfo
  {year} {2015}{\natexlab{b}})},\ \Eprint {http://arxiv.org/abs/1506.06767}
  {arXiv:1506.06767 [hep-ph]} \BibitemShut {NoStop}%
\bibitem [{\citenamefont {Schmitz}(2021)}]{Schmitz:2020syl}%
  \BibitemOpen
  \bibfield  {author} {\bibinfo {author} {\bibfnamefont {K.}~\bibnamefont
  {Schmitz}},\ }\href {\doibase 10.1007/JHEP01(2021)097} {\bibfield  {journal}
  {\bibinfo  {journal} {JHEP}\ }\textbf {\bibinfo {volume} {01}},\ \bibinfo
  {pages} {097} (\bibinfo {year} {2021})},\ \Eprint
  {http://arxiv.org/abs/2002.04615} {arXiv:2002.04615 [hep-ph]} \BibitemShut
  {NoStop}%
\bibitem [{\citenamefont {Dorsch}\ \emph {et~al.}(2017)\citenamefont {Dorsch},
  \citenamefont {Huber}, \citenamefont {Mimasu},\ and\ \citenamefont
  {No}}]{Dorsch:2017nza}%
  \BibitemOpen
  \bibfield  {author} {\bibinfo {author} {\bibfnamefont {G.~C.}\ \bibnamefont
  {Dorsch}}, \bibinfo {author} {\bibfnamefont {S.~J.}\ \bibnamefont {Huber}},
  \bibinfo {author} {\bibfnamefont {K.}~\bibnamefont {Mimasu}}, \ and\ \bibinfo
  {author} {\bibfnamefont {J.~M.}\ \bibnamefont {No}},\ }\href {\doibase
  10.1007/JHEP12(2017)086} {\bibfield  {journal} {\bibinfo  {journal} {JHEP}\
  }\textbf {\bibinfo {volume} {12}},\ \bibinfo {pages} {086} (\bibinfo {year}
  {2017})},\ \Eprint {http://arxiv.org/abs/1705.09186} {arXiv:1705.09186
  [hep-ph]} \BibitemShut {NoStop}%
\bibitem [{\citenamefont {Croon}\ \emph {et~al.}(2021)\citenamefont {Croon},
  \citenamefont {Gould}, \citenamefont {Schicho}, \citenamefont {Tenkanen},\
  and\ \citenamefont {White}}]{Croon:2020cgk}%
  \BibitemOpen
  \bibfield  {author} {\bibinfo {author} {\bibfnamefont {D.}~\bibnamefont
  {Croon}}, \bibinfo {author} {\bibfnamefont {O.}~\bibnamefont {Gould}},
  \bibinfo {author} {\bibfnamefont {P.}~\bibnamefont {Schicho}}, \bibinfo
  {author} {\bibfnamefont {T.~V.~I.}\ \bibnamefont {Tenkanen}}, \ and\ \bibinfo
  {author} {\bibfnamefont {G.}~\bibnamefont {White}},\ }\href {\doibase
  10.1007/JHEP04(2021)055} {\bibfield  {journal} {\bibinfo  {journal} {JHEP}\
  }\textbf {\bibinfo {volume} {04}},\ \bibinfo {pages} {055} (\bibinfo {year}
  {2021})},\ \Eprint {http://arxiv.org/abs/2009.10080} {arXiv:2009.10080
  [hep-ph]} \BibitemShut {NoStop}%
\bibitem [{\citenamefont {Ekstedt}\ \emph {et~al.}(2023)\citenamefont
  {Ekstedt}, \citenamefont {Schicho},\ and\ \citenamefont
  {Tenkanen}}]{Ekstedt:2022bff}%
  \BibitemOpen
  \bibfield  {author} {\bibinfo {author} {\bibfnamefont {A.}~\bibnamefont
  {Ekstedt}}, \bibinfo {author} {\bibfnamefont {P.}~\bibnamefont {Schicho}}, \
  and\ \bibinfo {author} {\bibfnamefont {T.~V.~I.}\ \bibnamefont {Tenkanen}},\
  }\href {\doibase 10.1016/j.cpc.2023.108725} {\bibfield  {journal} {\bibinfo
  {journal} {Comput. Phys. Commun.}\ }\textbf {\bibinfo {volume} {288}},\
  \bibinfo {pages} {108725} (\bibinfo {year} {2023})},\ \Eprint
  {http://arxiv.org/abs/2205.08815} {arXiv:2205.08815 [hep-ph]} \BibitemShut
  {NoStop}%
\bibitem [{\citenamefont {Schicho}\ \emph {et~al.}(2022)\citenamefont
  {Schicho}, \citenamefont {Tenkanen},\ and\ \citenamefont
  {White}}]{Schicho:2022wty}%
  \BibitemOpen
  \bibfield  {author} {\bibinfo {author} {\bibfnamefont {P.}~\bibnamefont
  {Schicho}}, \bibinfo {author} {\bibfnamefont {T.~V.~I.}\ \bibnamefont
  {Tenkanen}}, \ and\ \bibinfo {author} {\bibfnamefont {G.}~\bibnamefont
  {White}},\ }\href {\doibase 10.1007/JHEP11(2022)047} {\bibfield  {journal}
  {\bibinfo  {journal} {JHEP}\ }\textbf {\bibinfo {volume} {11}},\ \bibinfo
  {pages} {047} (\bibinfo {year} {2022})},\ \Eprint
  {http://arxiv.org/abs/2203.04284} {arXiv:2203.04284 [hep-ph]} \BibitemShut
  {NoStop}%
\bibitem [{\citenamefont {L\"ofgren}\ \emph {et~al.}(2023)\citenamefont
  {L\"ofgren}, \citenamefont {Ramsey-Musolf}, \citenamefont {Schicho},\ and\
  \citenamefont {Tenkanen}}]{Lofgren:2021ogg}%
  \BibitemOpen
  \bibfield  {author} {\bibinfo {author} {\bibfnamefont {J.}~\bibnamefont
  {L\"ofgren}}, \bibinfo {author} {\bibfnamefont {M.~J.}\ \bibnamefont
  {Ramsey-Musolf}}, \bibinfo {author} {\bibfnamefont {P.}~\bibnamefont
  {Schicho}}, \ and\ \bibinfo {author} {\bibfnamefont {T.~V.~I.}\ \bibnamefont
  {Tenkanen}},\ }\href {\doibase 10.1103/PhysRevLett.130.251801} {\bibfield
  {journal} {\bibinfo  {journal} {Phys. Rev. Lett.}\ }\textbf {\bibinfo
  {volume} {130}},\ \bibinfo {pages} {251801} (\bibinfo {year} {2023})},\
  \Eprint {http://arxiv.org/abs/2112.05472} {arXiv:2112.05472 [hep-ph]}
  \BibitemShut {NoStop}%
\end{thebibliography}%

\end{document}